\documentclass[%
reprint,
superscriptaddress,
 amsmath,amssymb,amsfonts,
 aps,
 pra,
 longbibliography
]{revtex4-2}

\newcommand{\ket}[1]{\ensuremath{\left| #1 \right>}}
\newcommand{\bra}[1]{\ensuremath{\left< #1 \right|}}

\newcommand{\cnj}[1]{{#1}^{\ast}}
\newcommand{\hcnj}[1]{{#1}^{\dagger}}

\newcommand{\avr}[1]{\ensuremath{\langle{#1}\rangle}}

\newcommand{\diag}{\mathop{\rm diag}\nolimits}

\renewcommand{\Re}{\mathop{\rm Re}\nolimits}
\renewcommand{\Im}{\mathop{\rm Im}\nolimits}
\newcommand{\Tr}{\mathop{\rm Tr}\nolimits}

\newcommand{\bs}[1]{\boldsymbol{#1}}
 \newcommand{\vc}[1]{\mathbf{#1}}

 \newcommand{\ind}[1]{\mathrm{#1}}

 \newcommand{\e}{\mathrm{e}}
 \newcommand{\ee}{\mathrm{e}}
  \newcommand{\dd}{\mathrm{d}}

\usepackage[english]{babel}
\usepackage[utf8x]{inputenc}
\usepackage[T1]{fontenc}

\usepackage{amsmath}
\usepackage[colorinlistoftodos]{todonotes}
\usepackage[colorlinks=true, allcolors=blue]{hyperref}
\usepackage{amssymb,amsmath,amsthm,mathtools,mathrsfs,hyperref}
\usepackage{graphicx}
\usepackage{multirow}
\usepackage{float}
\usepackage{subcaption}

\theoremstyle{definition}

\usepackage{dcolumn}
\usepackage{bm}

\begin{document}



\title{Lindblad dynamics of open multi-mode bosonic systems: Algebra of quadratic superoperators,
  exceptional points and speed of evolution}

\author{Andrei Gaidash}%
\email{andrei_gaidash@itmo.ru}
\affiliation{Laboratory of Quantum Processes and Measurements, ITMO University, 199034, 3b
  Kadetskaya Line, Saint Petersburg, Russia} 
\affiliation{Department of Mathematical Methods for Quantum
  Technologies, Steklov Mathematical Institute of Russian Academy of
  Sciences, 119991, 8 Gubkina St, Moscow, Russia}

\author{Alexei D. Kiselev}
\email{alexei.d.kiselev@gmail.com}
\affiliation{Laboratory of Quantum Processes and Measurements, ITMO University, 199034, 3b Kadetskaya Line, Saint Petersburg, Russia}
\affiliation{Leading Research Center "National Center for Quantum Internet", ITMO University, Birzhevaya Line,
  16, Saint Petersburg, 199034, Russia}


\author{Anton Kozubov}
\affiliation{Laboratory of Quantum Processes and Measurements, ITMO University, 199034, 3b
  Kadetskaya Line, Saint Petersburg, Russia} 
\affiliation{Department of Mathematical Methods for Quantum
  Technologies, Steklov Mathematical Institute of Russian Academy of
  Sciences, 119991, 8 Gubkina St, Moscow, Russia}

\author{George Miroshnichenko}
\affiliation{Waveguide Photonics Research Center, ITMO University, 197101, 49 Kronverksky Pr., Saint
  Petersburg, Russia} 
\affiliation{Institute <<High School of Engineering>>, ITMO University, 197101, 49 Kronverksky Pr.,
  Saint Petersburg, Russia}%

\date{\today}

\begin{abstract}
  Quantum dynamics of open continuous variable systems
  is a multifaceted rapidly evolving field of fundamental and technological significance.
An important example is a multimode photonic system coupled to
a Markovian bath
so that temporal evolution of
the density matrix  is governed by the master equation of the Lindblad form.
Theoretical analysis of the
Lindblad dynamics complicated by intermode couplings
such as dynamical (coherent) and environment-mediated (incoherent)
  interactions between the modes
can be rather involved even for exactly soluble models.
Our algebraic technique
uses the Lie algebra of quadratic combinations of left and
right superoperators to simplify both quantitative and qualitative analysis of
intermode coupling-induced effects in multi-mode systems by
eliminating jump superoperators and
reducing the Lindblad equation to
the form governed by the effective non-Hermitian Hamiltonian, $\hat{H}_{\ind{eff}}$.
We show that the spectral properties of the thermal bath Liouvillian
are determined by the effective temperature independent Hamiltonan 
and employ  jump-eliminating transformations for computing
its eigenvalues and eigenoperators. 
It is found that the Liouvillian exceptional points (EPs) are the points in the parameter space where
the matrix,  $H$, associated with $\hat{H}_{\ind{eff}}$ is nondiagonalizable.
For the case of photonic polarization modes
modeled as a bimodal bosonic system, we describe the geometry of EPs
  and apply our method to study how
  the intermode-couplings-induced effects such as EPs and temperature influence
  the dynamics of the speed of evolution that governs
  the quantum speed limit times of polarization qubit states.  
\end{abstract}

\maketitle


\section{Introduction}
\label{sec:intro}

The importance of the problems related to
quantum dynamics of open systems
for quantum technologies
such as
quantum communications and quantum computations
cannot be overestimated.
The theory of quantum open systems
is essential to
the understanding of
the processes  that underlie
transfer and storage of quantum information
and are
generally described in terms of
completely positive, hermiticity and trace-preserving maps known as
  the quantum channels
(see,
e.g.~\cite{Stinespring:procams:1955,Choi:linalg:1975,Kraus:bk:1983,Holevo:problemy:1972,Caruso:rmp:2014}
for analysis of the mathematical structures
related to the quantum channels).

An alternative and a widely used
master equation  approach to quantum systems
that evolve in time
interacting with an environment
assumes that temporal evolution 
of the reduced density matrix, $\hat{\rho}$,
is governed by the equation of motion of the general form
\begin{align}
  \label{eq:master-gen}
  \frac{\partial\hat{\rho}}{\partial t}=\hat{\mathcal L}\hat{\rho},
\end{align}
where $\hat{\mathcal L}$ is the Liouvillian superoperator
(a linear mapping acting on the vector space of linear operators).
When the evolution is unitary, so that
$\hat{\rho}(t)=\hat{U}(t)\hat{\rho}(0)\hcnj{\hat{U}}(t)$,
where $\hat{U}(t)$ is the unitary evolution operator
and the dagger denotes Hermitian conjugation,
the Liouvillian, $\hat{\mathcal L}=\hat{\mathcal L}_{H}$,
is determined by the generator of $\hat{U}$ giving
the Hermitian Hamiltonian, $\hat{H}$, as follows
\begin{align}
  \label{eq:L-unitary}
  \hat{\mathcal L}_H\hat{\rho}=-i[\hat{H},\hat{\rho}],
  \quad
  \hat{\mathcal L}_H=-i(\overleftarrow{\hat{H}}-\overrightarrow{\hat{H}}),
\end{align}
where $[\hat{A},\hat{B}]=\hat{A}\hat{B}-\hat{B}\hat{A}$
is the commutator and the left (right) arrow over an operator
$\overleftarrow{\hat{A}}$ ($\overrightarrow{\hat{A}}$)
indicates the left-hand (right-hand) action
superoperator:
$\overleftarrow{\hat{A}}:\:\hat{\rho}\mapsto{\hat{A}}\hat{\rho}\equiv\overleftarrow{\hat{A}}\hat{\rho}$
($\overrightarrow{\hat{A}}:\hat{\rho}\mapsto\hat{\rho}{\hat{A}}\equiv\overrightarrow{\hat{A}}\hat{\rho}$).
For the general case of nonunitary dynamics,
the Liouvillian is supplemented by the dissipative terms
describing loss of energy and information into the environment.

The literature on a variety of master equations
derived using different assumptions and
approximations
is quite extensive~\cite{Carmichael:bk1:2002,Omnes:pra:1997,Breuer:bk:2002,Rivas:bk:2012,Rivas:repprogphys:2014,deVega:rmp:2017}.
In particular,
under certain rather general assumptions,
the dissipative terms are captured by the so-called Lindblad dissipators
leading to 
the well-known Lindblad
form of master equations~\cite{Kossakowski:repmathphys:1972,Lindblad:commathphys:1976,Gorini:jmathphys:1976}
(the Lindblad master equations are alternatively called
the Gorini-Kossakowski-Sudarshan-Lindblad (GKSL) equations).

The general physics behind the Lindblad-type master equations
has been extensively discussed
in the literature
(see, e.g.,
Refs.~\cite{Pearle:ejp:2012,Albash:njp:2012,McCauley:npj:2020,Manzano:aip:2020}).
Interestingly, in the framework of measurement theory,
the Lindblad dissipators can be interpreted
as the terms that determine the temporal evolution of a system continuously
probed by the environment.
They can be split into two parts:
(a)~the continuous non-unitary dissipation terms
of the form $\overleftarrow{\hat{\Gamma}}+\overrightarrow{\hat{\Gamma}}$
that in combination with the unitary Liouvillian~\eqref{eq:L-unitary}
give the generalized commutator
$\overleftarrow{\hat{H}_{\ind{eff}}}-\overrightarrow{\hat{H}_{\ind{eff}}^{\dagger}}$
extended to the case of a non-Hermitian effective Hamiltonian
where $\hat{H}_{\ind{eff}}=\hat{H}+i \hat{\Gamma}$;
and (b)~the quantum jump part
that contains the so-called jump terms of the form
$\hat{\mathcal L}_J=\overleftarrow{\hat{J}}\overrightarrow{\hcnj{\hat{J}}}$,
where $\hat{J}$ is the jump operator,
describing the effects of the measurements performed by the
environment
(alternatively, such terms are sometimes called the recycling terms).

  Note that this interpretation implies that the environment
can be replaced with a measuring device where measurement results
are not recorded.
When evolution is conditioned on the results of measurements
performed by a perfect device,
the dynamics can be formulated in terms of pure states
describing
the so-called quantum trajectories
(details on the quantum trajectory approach can be found, e.g., in
reviews~\cite{Carmichael:bk:1993,Plenio:rmp:1998}),
so that, for a sufficiently large sample, averaging over the trajectories
will produce the expectation values
obtained using the density matrix $\hat{\rho}$.
The latter is at the heart of a variety of
mathematically equivalent computational approaches
that differ in numerical implementations
such as the quantum jump approach,
the Monte Carlo wavefunction, and the stochastic sampling
methods
(see, e.g. Refs.~\cite{Daley:advph:2014,Verstraelen:appsci:2018,Verstraelen:phd:2020,Verstraelen:prxq:2023}).

In this paper, the special case of open continuous-variable systems
will be our primary concern.
These systems are
of fundamental importance for an understanding
of the dissipative dynamics of the modes of quantum bosonic fields, such as
quantized light.
Whole fields of science and technology, such as the quantum key distribution~\cite{bennett1984quantum,
  pirandola2020advances}, quantum teleportation~\cite{pirandola2015advances},
quantum sensing~\cite{Degen:rmp:2017,pirandola2018advances}, and others have emerged due to the unique properties of quantized light,
unmatched to those of classical optics. In some way or another, all pursue a goal: reducing the
effects of decoherence. So, one of the aspects of the theory of  open quantum systems is
understanding the physics related to the decoherence effects and capitalize the knowledge in application
areas. 

For example, the
quantum dynamics of the polarization states of the photonic modes propagating in optical fibers
can be modeled within the framework of the Lindblad dynamics of a two-mode bosonic
system~\cite{miroshnichenko2018decoherence,kozubov2019quantum,gaidash2020dissipative,gaidash2021quantum}.
A family of similar models has been
the subject of intense
studies~\cite{Hiroshima:pra:2001,Serafini:pra:2004,Castro:pla:2008,Chou:pre:2008,Paz:prl:2008,Paz:pra:2009,Barbosa:pra:2011,Figueiredo:physa:2016,Linowski:pra:2020,Vendromin:pra:2021,Kiselev:entropy:2021}
on  the dynamics of entanglement
in open systems of two coupled oscillators.

More recent examples include 
modeling quantum scattering from a lossy bianisotropic
metasurface as a Lindblad master equation propagating the photon-moment matrix
in fictitious time~\cite{Wong:pra:2022} and
applications of the Lindblad dynamics approach
to non-Hermitian systems with exceptional points (EPs)
such as waveguides with parity-time ($\mathscr {PT}$), anti-$\mathscr{PT}$ ,
and Floquet $\mathscr{PT}$
symmetries~\cite{Minganti:pra:2019,Arkhipov:pra:2020,Minganti:pra:2022,Nakanishi:pra:2022}.

EPs defined as degeneracies of non-Hermitian Hamiltonians
arising when both eigenvalues and eigenmodes coalesce
have been extensively studied in the context
of non-Hermitian systems with  $\mathscr{PT}$ symmetry
(reviews on the physics of $\mathscr{PT}$ symmetric photonic systems
can be found
in~\cite{Ozdemir:nature:2019,Parto:nanoph:2021,Wang:jopt:2021}).
In such systems, EPs appear as
transition points that separate
two spectrally and dynamically distinct regions:
a spectral region with entirely
real eigenvalues (the so-called $\mathscr{PT}$ unbroken
region) and one without
(the so-called $\mathscr{PT}$ broken region).

EPs also exist in systems without the $\mathscr{PT}$ symmetry, and
their singular nature manifests itself  in a variety of interesting phenomena,
including
single-mode lasing~\cite{Feng:sci:2014},
enhancement of spontaneous emission~\cite{Pick:optexp:2017,Zhou:aplq:2024},
enhanced entanglement generation~\cite{Li:prl:2023,Han:prl:2023,Tang:lsa:2024} 
and
strong sensitivity to external
perturbations
that lies at the heart of EP-enhanced
sensing~\cite{Wiersig:prl:2014,Wiersig:pra:2016,Wiersig:pra:2020,Zhang:prl:2019,Anderson:prappl:2023,Waghela:avs:2024}.





In Refs.~\cite{Am-Shallem:njp:2015,Minganti:pra:2019},
the concept of EPs is extended to the case of quantum Liouvillians.
Such Liouvillian EPs (LEPs) are determined by the spectral properties of
the Liouvillian superoperator.
In addition to finite-dimensional qudit systems
(see Ref.~\cite{Sun:aapps:2024} for a recent review focused on LEPs in atomic systems),
LEPs have been explored in a  number of continuous-variable
systems~\cite{Arkhipov:pra:2020,Minganti:pra:2022,Nakanishi:pra:2022,Chimczak:scirep:2023}.
Dynamics of these systems is governed by
thermal bath multimode bosonic
Lindbladians and, in general,
can be very complex
due to interactions between different modes
(dynamical (coherent) and
environment-mediated
(incoherent)
intermode
couplings are discussed, e.g., in~\cite{Hackenbroich:pra:2003,gaidash2020dissipative,gaidash2021quantum,Franke:prl:2019,Kiselev:entropy:2021}).
The above discussed physical phenomena near EPs and LEPs
exemplify important modern problems that cannot be understood
without theoretical analysis
of the effects induced by intermode couplings.
In this paper, our goal is to develop an algebraic method
that provides an efficient tool for solving the spectral problem for
the Liouvillian superoperators of multi-mode systems.

It should be emphasized that,
even though the dissipative dynamics
can be simulated numerically
using software packages such as the Python package
QuTiP~\cite{Johansson:cpc:2012},
there are several reasons why a comprehensive study would benefit
from analytical approaches:
(a)~owing to the complexity of interactions,
it could be challenging to understand how the multiple parameters influence dynamical
    regimes and the related effects;
    (b)~contributions from infinite-dimensional density
          matrices can affect the accuracy of numerical results using finite dimensional
    approximations, whereas
continuous-variable exact calculations can provide an advantage with respect to imposing a
cutoff in the Hilbert space~\cite{Stornati:arxiv:2023}; 
(c)~under certain conditions (such as the presence of EPs), numerical
          calculations may not converge  and produce accurate results.
          So, in what follows, the algebraic method that underlies
          a systematic analytical approach
          to the eigenspectrum problem of thermal bath Lindblandians
          will be our primary concern.
          
The bulk of mathematical
techniques developed for analysis of bosonic systems are mostly applicable to
the single-mode Lindblad
equation~\cite{Briegel:pra:1993,Briegel:pra:1994,Arevalo:qsopt:1998,Klimov:joptb:2003,Lu:pra:2003,Tay:jmp:2008,Ban:jmo:2009,Honda:jmp:2010,Tay:physica:2020,Shishkov:pra:2020}
and cannot be employed to treat its multi-mode generalizations
(the most general form of the multi-mode Lindblad
equation is described, e.g.,
in~\cite{Benatti:jpa:2006,Prosen:jpa:2010,Kiselev:symmetry:2021}).


More specifically,
in
Refs.~\cite{Briegel:pra:1993,Briegel:pra:1994,Barnett:jmo:2000,Tay:jmp:2008,Ban:jmo:2009,Honda:jmp:2010,Tay:physica:2020},
analytical solutions to
the spectral problem
were derived using different methods.
Algebraic approaches
have been the subject of studies
on Lie algebra methods~\cite{Ban:pra:1993,Arevalo:qsopt:1998,Abdalla:qsoptics:1998,Lu:pra:2003}
and
the symmetries of Lindblad equations~\cite{Tay:pra:2007,Albert:pra:2014}.
Alternative algebraic methods were put forward in
Refs.~\cite{Klimov:joptb:2003,Nakazato:pra:2006,Tay:physa:2017}.

In the recent paper~\cite{Teuber:pra:2020},
Fock-like eigenstates of Lindbladian
are constructed using Lie algebras induced by
the master equation for a linear chain of coupled harmonic oscillators.
In Refs.~\cite{McDonald:prl:2022,Nakanishi:pra:2022},
multimode systems
were studied using the formalism of
bosonic third quantization
that can be regarded as canonical quantization in the Liouville space~\cite{Prosen:njp:2008,Prosen:jpa:2010}.

In this paper, we present an algebraic method
based on the algebra of quadratic superoperators associated with matrices
that simplifies the analysis of the multi-mode systems with complicated inter-mode couplings.
In general, this approach provides an alternative prospect on the Lindblad dynamics of continuous variable
optical quantum systems emphasizing the relationship between
the algebraic structure of the steady state and the superoperators that enter the jump-eliminating transformations. 
The paper is organized as follows.

In Sec.~\ref{sec:algebra}
after introducing multi-mode Liouvillian
and related superoperators,
we generalize the Jordan-Wigner approach to the case of superoperators
and examine
the algebraic structure of quadratic superoperators.
The solution of the spectral problem
is presented in Sec.~\ref{sec:spectral-problem}
where the algebraic relations are applied to
eliminate the jump (cycling) terms
and transform
the Liouvillian into the so-called diagonalized form.
This form is determined by
the effective non-Hermitian Hamiltonian
which, under certain conditions,
can be further simplified by introducing
the representation of normal modes.

In Sec.~\ref{sec:applications}
we apply our algebraic technique
to bimodal photonic (bosonic) systems
whose effective Hamiltonians
are characterized by
the frequency and relaxation vectors
describing
the coherent and incoherent couplings
between the modes, respectively.
We show that, at LEP,
where the dynamics experiences a slow-down,
these vectors are orthogonal, and their lengths are equal.
As an illustration, the formalism of quadratic superoperators
is employed to study the speed of evolution of polarization qubit
in the low-temperature regime, where
the mean number of thermal photons, $n_T$, is small.
This speed can be regarded as
a measure sensitive to fine details of the Lindblad dynamics.
It is evaluated at varying coupling parameters and temperatures.
Section~\ref{sec:conclusion} concludes the paper.
The technical details are relegated to the
Appendices~\ref{app:Riccati}--~\ref{appder}.

\section{Algebraic approach}
\label{sec:algebra}

In this section, we show that
the Jordan-Schwinger approach
extended to the case of left and right superoperators
provides an efficient and natural way to describe
the algebraic structure of superoperators that
enter bosonic GKSL equations.

\subsection{Model}\label{subsec:problem}

Lindblad dynamics of a $N$-mode bosonic system
interacting with a thermal environment
is governed by the Lindblad (GKSL) equation
of the following form
(our notations are close to those used in Ref.~\cite{gaidash2021quantum}):
\begin{align}
  &
    \frac{\partial \hat{\rho}}{\partial
      t}=- i\sum_{n,m}^N\Omega_{n m}[\hat{a}_n^{\dagger}\hat{a}_m,\hat{\rho}]
    \notag
  \\
  &
    -\sum_{n,m}^N
    \bigl\{
    \Gamma_{n m}^{(-)}
    (\hat{a}_n^{\dagger}\hat{a}_m\hat{\rho}+\hat{\rho}\hat{a}_n^{\dagger}\hat{a}_m-2\hat{a}_m\hat{\rho}\hat{a}_n^{\dagger})
    \notag
  \\
  &
    +\Gamma_{n m}^{(+)}(\hat{a}_m\hat{a}_n^{\dagger}\hat{\rho}+\hat{\rho}\hat{a}_m\hat{a}_n^{\dagger}-2\hat{a}_n^{\dagger}\hat{\rho}\hat{a}_m)
    \bigr\},
    \label{eq:L-rho}
  \\
  &
    \label{eq:Gamma_pm}
    \Gamma_{\pm}=\gamma_{\pm}\Gamma,\quad
    \Gamma=\Gamma_{-}-\Gamma_{+},\quad
    \gamma_{+}=\gamma_{-}-1=n_T,
\end{align}
where
the dagger stands for Hermitian conjugation,
$\hat{a}_n$ ($\hat{a}^\dagger_n$) is the annihilation (creation) operator of $n$th mode
and $n_T$ is the mean number of thermal photons;
the effects of coherent (dynamical)
intermode couplings are introduced through
the frequency matrix
$\Omega=\hcnj{\Omega}$,
whereas
the
elements of the positive definite relaxation matrices
$\Gamma_{\pm}>0$
and $\Gamma=\Gamma_{-}-\Gamma_{+}>0$
give the coupling constants of
the incoherent  (environment-mediated) interaction
between the bosonic modes.

Our task now is
to put the thermal bath multimode Liouvillian
in a form suitable for subsequent algebraic analysis.
To this end,
we shall generalize considerations presented
in Ref.~\cite{vadeuiko2000diagonal,gaidash2020dissipative} to
the case of  multimode systems
and introduce quadratic combinations
of left and right superoperators
of the following form:
\begin{subequations}
  \label{eq:NM_operators}
\begin{align}
  &
    \label{eq:Nm_nm}
    \hat{\mathcal N}^{(-)}_{nm}=\overleftarrow{\hat{a}^\dagger_n \hat{a}_m}-\overrightarrow{\hat{a}^\dagger_n
    \hat{a}_m},
    \\
     &
    \label{eq:K0_nm}
  \hat{\mathcal K}_{nm}^{(0)}=\frac{1}{2}\left(\overleftarrow{\hat{a}^\dagger_n\hat{a}_m}
       +\overrightarrow{\hat{a}_m\hat{a}^\dagger_n}\right).
  \\
  &
    \label{eq:Kpm_nm}
    \hat{\mathcal K}_{nm}^{(+)}=\overleftarrow{\hat{a}^\dagger_n} \overrightarrow{\hat{a}_m},
    \quad
    \hat{\mathcal K}_{nm}^{(-)}=\overleftarrow{\hat{a}_m}\overrightarrow{\hat{a}^\dagger_n}.
  \end{align}
\end{subequations}
The Liouvillian
can now be expressed in terms of
the superoperators~\eqref{eq:NM_operators}
as follows
\begin{align}
  &
    \label{L1}
    \hat{\mathcal L}=\sum_{n,m}^N\Bigl[-i\Omega_{nm}\hat{\mathcal N}^{(-)}_{nm}
    + 2 \bigl\{\Gamma_{nm}^{(0)}\hat{\mathcal K}_{nm}^{(0)}
    \notag
  \\
  &
    +\Gamma_{nm}^{(+)}\hat{\mathcal K}_{nm}^{(+)}
    +\Gamma_{nm}^{(-)}\hat{\mathcal K}_{nm}^{(-)}
    \bigr\}
+\delta_{nm}\Gamma_{nm}
    \Bigr],
  \\
  &
    \label{eq:Gamma_0}
    \Gamma_0=-(\Gamma_{+}+\Gamma_{-})=\gamma_0\Gamma,
    \quad \gamma_0=-2n_T-1,
\end{align}
where 
$\delta_{nm}$ is the Kronecker symbol.
Note that
the terms proportional to the jump superoperators
$\hat{\mathcal K}_{nm}^{(+)}$ and $\hat{\mathcal K}_{nm}^{(-)}$
represent
the quantum jump part of $\hat{\mathcal L}$,
whereas 
the effective Hamiltonian
is given by
\begin{align}
  &
  \label{eq:Heff_0}
  \hat{H}_{\ind{eff}}^{(0)}=
  \sum_{n,m}^N\left[\Omega_{n m}+i\Gamma_{nm}^{(0)}
  \right]\hat{a}_n^{\dagger}\hat{a}_m
  -i \Tr{\Gamma_{+}}.
\end{align}

The formula~\eqref{L1}
is the starting point of our analysis.
It gives the multimode Liouvillian
written as a linear combination
of the superoperators:
$\hat{\mathcal N}^{(-)}_{nm}$, $\hat{\mathcal K}_{nm}^{(0)}$ and
$\hat{\mathcal K}_{nm}^{(\pm)}$ given by Eq.~\eqref{eq:NM_operators}.
An important point is that these superoperators
generate a Lie algebra.
However,
owing to a number of factors, such as large dimension,
the structure of this algebra is prohibitively complex.
In subsequent sections,
we describe the approach
that allows us to
clarify an overall picture behind the complicated algebraic structure
by greatly reducing the number of
commutation relations. 

\subsection{Jordan map}
\label{subsec:Jordan}

Now we briefly remind the reader
about the
well-known Jordan-Schwinger approach that uses the Jordan map
(more details can be found in, e.g., Chapter~5 of the book~\cite{Biedenharn:bk:1984})
given by
\begin{align}
  \label{eq:Jordan}
  A\mapsto \hat{J}_{A}=\sum_{n,m=1}^{N}A_{nm}\hat{a}^\dagger_n\hat{a}_m,
\end{align}
where $A$ is the $N\times N$ matrix, 
to introduce quadratic boson operators associated with matrices.
It is rather straightforward to check that these operators meet the commutation relations
\begin{align}
  \label{eq:Jordan-comm-rel}
  [\hat{J}_{A},\hat{J}_{B}]=\hat{J}_{[A,B]}
\end{align}
and enjoy a number of useful properties
including the indentities
\begin{subequations}
  \label{eq:Jordan-ident}
\begin{align}
  &
  \label{eq:Jordan-ident-1}
  e^{\hat{J}_V}\hat{J}_{A} e^{-\hat{J}_V}=\hat{J}_{A'},
  \quad
  A'=e^V A e^{-V},
  \\
  &
  \label{eq:Jordan-ident-2}
  e^{\hat{J}_V}\hat{a}^{\dagger}_ie^{-\hat{J}_V}=
  \sum_{j=1}^N (e^V)_{ji}\hat{a}^{\dagger}_j,
  \\
  &
    \label{eq:Jordan-ident-3}
  e^{\hat{J}_V}\hat{a}_ie^{-\hat{J}_V}=
  \sum_{j=1}^N (e^{-V})_{ij}\hat{a}_j
\end{align}
\end{subequations}
that will be used in our subsequent calculations.

Our next step is to introduce the algebra of superoperators which
bears some resemblance to
the above algebra with
the commutation relations~\eqref{eq:Jordan-comm-rel}.
The idea is to use bilinear combinations
of the superoperators so that the commutation relations appear to be partly translated into the
change of the matrices of the coefficients.

Now we move on to discussing the basic algebraic
properties, whereas
the spectral problem and
other applications will be considered later on.

\subsection{Superoperators associated with matrices and Liouvillian}
\label{subsec:super-assoc-with}

Let us introduce the following notation:
\begin{align}
\label{eq:P_A}
    \hat{\mathcal{P}}_{A}
    \equiv \sum_{n,m} A_{nm} \hat{\mathcal{P}}_{nm},
\end{align}
where $\hat{\mathcal{P}}_{nm}$ is an arbitrary (presumably quadratic) superoperator labeled with two
indices and $A_{nm}$ is the element of a complex-valued matrix $A$. Then, according to the above
notation, the superoperators have the following general properties:
\begin{gather}
    \hat{\mathcal{P}}_{A}+\hat{\mathcal{Q}}_{A}=(\hat{\mathcal{P}}+\hat{\mathcal{Q}})_{A},\\
    \hat{\mathcal{P}}_{A}+\hat{\mathcal{P}}_{B}=\hat{\mathcal{P}}_{A+B},\\
    c\cdot\hat{\mathcal{P}}_{A}=\hat{\mathcal{P}}_{c\cdot A}.
\end{gather}
Another useful property is that, when the operators meet the commutation relations of the form: 
\begin{gather}
    [\hat{\mathcal{P}}_{ij},\hat{\mathcal{Q}}_{nm}]=\hat{\mathcal{S}}_{im}\delta_{jn}\pm\hat{\mathcal{S}}_{jn}\delta_{im},\label{prop1}
\end{gather}
where $[A,B]=AB-BA$ is the commutator, we have the following identity:
\begin{gather}
  [\hat{\mathcal{P}}_{A},\hat{\mathcal{Q}}_{B}]=\hat{\mathcal{S}}_{AB\pm BA},
  \label{prop2}
\end{gather}
which is analogous to the key property of the Jordan-Schwinger map given by Eq.~\eqref{eq:Jordan-comm-rel}.
However, as we will
see in the next section, the algebraic structure of superoperators associated with matrices has a
number of important distinctive peculiarities.

By using these superoperators,
the Liouvillian~\eqref{L1}
may alternatively be rewritten as follows:
\begin{align}
  &
  \label{eq:L-gen-form}
    \hat{\mathcal L}=-i\hat{\mathcal{N}}^{(-)}_{\Omega}+
    2 \bigl(
    \hat{\mathcal{K}}^{(0)}_{\Gamma_0}
    +
    \hat{\mathcal{K}}^{(+)}_{\Gamma_{+}}+\hat{\mathcal{K}}^{(-)}_{\Gamma_{-}}
    \bigr)
    +\Tr\Gamma\,
    \hat{\mathcal{I}},
\end{align}
where
$\hat{\mathcal{I}}$ is the identity superoperator.
Note that the conjugate (adjoint) of the Liouvillian superoperator,
$\hat{\mathcal L}^{\sharp}$, defined through the relation
\begin{align}
  \label{eq:L_cnj}
  \mathrm{Tr}\{\hat{A}\hat{\mathcal L}(\hat{\rho})\}=\mathrm{Tr}\{\hat{\mathcal L}^{\sharp}(\hat{A})\hat{\rho}\}
\end{align}
is also of the form~\eqref{eq:L-gen-form}
\begin{align}
  \label{eq:L-sharp-gen}
  \hat{\mathcal L}^{\sharp}=i\hat{\mathcal{N}}^{(-)}_{\Omega}+
    2 \bigl(
    \hat{\mathcal{K}}^{(0)}_{\Gamma_0}
    +
    \hat{\mathcal{K}}^{(+)}_{\Gamma_{-}}+\hat{\mathcal{K}}^{(-)}_{\Gamma_{+}}
    \bigr)
    +\Tr\Gamma\,
    \hat{\mathcal{I}},
\end{align}
where $\Omega$ is changed to $-\Omega$
and the jump parameters $\gamma_{+}$ and $\gamma_{-}$
are interchanged: $\gamma_{\pm}\to\gamma_{\mp}$.
Now we discuss the basic commutation
properties of the superoperators that enter the above expression for the Liouvillian.

\subsection{Commutation relations}
\label{subsec:comm-relat}

In this section, our goal is to clarify the algebraic structure of the superoperators.  Since this
structure is determined by commutation relations, we begin with the case of superoperator
$\hat{\mathcal N}^{(-)}_{nm}$, $\hat{\mathcal K}_{nm}^{(+)}$, $\hat{\mathcal K}_{nm}^{(-)}$, and $\hat{\mathcal K}_{nm}^{(0)}$ and
deduce the relations for superoperators of the form~\eqref{eq:P_A} associated with matrices.

It is rather straightforward to obtain 
the relations
\begin{subequations}
  \label{eq:alg-comm-rels-0}
\begin{align}
  &
    [\hat{\mathcal K}^{(0)}_{ij},\hat{\mathcal K}^{(\pm)}_{nm}]=\pm\frac12(\hat{\mathcal
      K}^{(\pm)}_{im}\delta_{jn}+\hat{\mathcal K}^{(\pm)}_{jn}\delta_{im}),
  \\
  &
    [\hat{\mathcal K}^{(-)}_{ij},\hat{\mathcal K}^{(+)}_{nm}]=\hat{\mathcal
      K}^{(0)}_{im}\delta_{jn}+\hat{\mathcal K}^{(0)}_{jn}\delta_{im}-
    \notag
  \\
  &
    -\frac12\big(\hat{\mathcal N}^{(-)}_{im}\delta_{jn}-\hat{\mathcal N}^{(-)}_{jn}\delta_{im}\big),
  \\
  &
    [\hat{\mathcal N}^{(-)}_{ij}, \hat{\mathcal K}^{(s)}_{nm}]=\hat{\mathcal
      K}^{(s)}_{im}\delta_{jn}-\hat{\mathcal K}^{(s)}_{jn}\delta_{im},
    \quad
    s\in\{0,\pm\}
\end{align}
\end{subequations}
and employ the properties given by Eqs.~\eqref{prop1} and~\eqref{prop2} to derive
the following commutation relations
\begin{subequations}
  \label{eq:alg-comm-rels}
\begin{align}
  \label{algebra1}
  &
    [\hat{\mathcal{K}}^{(0)}_{A}, \hat{\mathcal{K}}^{(\pm)}_{B}]
  =\pm\hat{\mathcal{K}}^{(\pm)}_{\frac12\{A,B\}},
  \\
  \label{algebra2}
  &
    [\hat{\mathcal{K}}^{(-)}_{A}, \hat{\mathcal{K}}^{(+)}_{B}]
  =\hat{\mathcal{K}}^{(0)}_{\{A,B\}}-\hat{\mathcal{N}}^{(-)}_{\frac12[A,B]},
  \\
  \label{algebra3}
  &
  [\hat{\mathcal{N}}^{(-)}_{A}, \hat{\mathcal{K}}^{(s)}_{B}]=\hat{\mathcal{K}}^{(s)}_{[A,B]},
  \\
  \label{algebra01}
  &
     [\hat{\mathcal{K}}^{(0)}_{A}, \hat{\mathcal{K}}^{(0)}_{B}]
  =\hat{\mathcal{N}}^{(-)}_{\frac14[A,B]},
  \\
  \label{algebra01-1}
  &
     [\hat{\mathcal{N}}^{(-)}_{A}, \hat{\mathcal{N}}^{(-)}_{B}]
  =\hat{\mathcal{N}}^{(-)}_{[A,B]},
  \\
  \label{algebra4}
  &
  [\hat{\mathcal{K}}^{(\pm)}_{A}, \hat{\mathcal{K}}^{(\pm)}_{B}]=0,
\end{align}
\end{subequations}
where $\{A,B\}=AB+BA$ stands for the anticommutator.

Formulas~\eqref{eq:alg-comm-rels} describe the algebraic structure that somewhat
resembles $\mathfrak{su}(1,1)$ Lie algebra.
A direct sum of such algebras
is exactly the algebra corresponding to
the limiting case of noninteracting modes, where all the matrices that enter
Eq.~\eqref{eq:L-gen-form} are diagonal
($\{\hat{\mathcal K}^{(0)}_{nn},\hat{\mathcal K}^{(+)}_{nn},\hat{\mathcal K}^{(-)}_{nn}\}$
are the generators of $\mathfrak{su}(1,1)$ Lie algebra).
However, the effects of intermode couplings cannot generally be neglected, and thus the
matrices $\Omega$ and $\Gamma$ are not necessarily diagonal
(both are Hermitian and $\Gamma>0$). So, we focus our attention on this general case.

\section{Spectral problem}
\label{sec:spectral-problem}

In this section, we shall use the above algebraic relations
to solve the spectral problem for the Liouvillian~\eqref{L1}.
Our analysis involves two steps:
(a)~applying a transformation that casts the Liouvillian
into the diagonalized form which is solely expressed in terms of the
effective non-Hermitian Hamiltonian
and does not contain
terms proportional to the jump superoperators  $\hat{\mathcal K}_{nm}^{(\pm)}$;
and (b)~simplifying the effective Hamiltonian
so as to eliminate the inter-mode interactions.

\subsection{Jump-eliminating transformations}
\label{subsec:Ldiag}

Similarly to Ref.~\cite{gaidash2020dissipative},
we begin with algebraic identities for exponentiated
superoperators acting by similarity (adjoint action).  
By using the commutation relations~\eqref{eq:alg-comm-rels},
we have
\begin{subequations}
  \label{eq:transform}
\begin{align}
    e^{\hat{\mathcal{K}}^{(\pm)}_{B}} \hat{\mathcal{N}}^{(-)}_{A} e^{-\hat{\mathcal{K}}^{(\pm)}_{B}}
  &= \hat{\mathcal{N}}^{(-)}_{A} - \hat{\mathcal{K}}^{(\pm)}_{[A,B]},
    \label{orth1}
  \\
    e^{\hat{\mathcal{K}}^{(\pm)}_{B}} \hat{\mathcal{K}}^{(0)}_{A} e^{-\hat{\mathcal{K}}^{(\pm)}_{B}}
  &= \hat{\mathcal{K}}^{(0)}_{A}\mp\hat{\mathcal{K}}^{(\pm)}_{\frac12\{A,B\}},
  \\ 
    e^{\hat{\mathcal{K}}^{(\pm)}_{B}} \hat{\mathcal{K}}^{(\mp)}_{A} 
  e^{-\hat{\mathcal{K}}^{(\pm)}_{B}}
  &= \hat{\mathcal{K}}^{(\mp)}_{A}\mp\hat{\mathcal{K}}^{(0)}_{\{A,B\}} 
    \notag
    \\
    &
    +\hat{\mathcal{N}}^{(-)}_{\frac12[A,B]}+\hat{\mathcal{K}}^{(\pm)}_{BAB}\label{orth2}.
\end{align}
\end{subequations}
From these identities,
the transformations determined by the jump operators,
$\hat{\mathcal{K}}^{(\pm)}_{B}$,
modify a linear combination
of superoperators that enter the Liouvillian~\eqref{eq:L-gen-form}
as follows:
\begin{subequations}
  \label{eq:L-via-Kpm}
\begin{align}
  &
  \label{eq:L-prime}
    \hat{\mathcal L}=-i\hat{\mathcal{N}}^{(-)}_{\Omega}+2
    \sum_{\alpha\in\{0,\pm\}}\hat{\mathcal{K}}^{(\alpha)}_{\Gamma_{\alpha}}+
    \Tr\Gamma\,\hat{\mathcal{I}}
    \notag
  \\
  &
    \mapsto 
    \hat{\mathcal L}'=
    -i\hat{\mathcal{N}}^{(-)}_{\Omega'}+2\sum_{\alpha\in\{0,\pm\}}\hat{\mathcal{K}}^{(\alpha)}_{\Gamma'_{\alpha}}+
    \Tr\Gamma\,\hat{\mathcal{I}}
    \notag
  \\
  &
= e^{\hat{\mathcal{K}}^{(\nu)}_{B}}\hat{\mathcal
    L}e^{-\hat{\mathcal{K}}^{(\nu)}_{B}},
    \quad \nu\in\{+,-\},
  \\
  &
    \label{eq:OmG0-prime}
    \Omega'= \Omega+i [\Gamma_{-\nu},B],
  \quad
    \Gamma_{0}' = \Gamma_0-\nu\{\Gamma_{-\nu},B\},
  \\
  &
    \label{eq:Gamma_nu-prime}
    \Gamma_{\nu}' = \Gamma_{\nu}+\frac{i}{2}[\Omega,B]-
    \frac{\nu}{2}\{\Gamma_0,B\}+B\Gamma_{-\nu}B,
  \\
  &
    \label{eq:Gamma_mnu-prime}
    \Gamma_{-\nu}' = \Gamma_{-\nu},
\end{align}
\end{subequations}
where the special case of thermal bath Liouvillians
is described by
the superoperator~\eqref{eq:L-gen-form}
with
the matrices $\Gamma_{\alpha}=\gamma_{\alpha}\Gamma$
and the coefficients $\gamma_{\alpha}$ given
by Eqs.~\eqref{eq:Gamma_pm} and~\eqref{eq:Gamma_0}.

Now it is our task to
simplify the Liouvillian by eliminating
contributions coming from jump operators
with the help of compositions of
the above similarity transformations~\eqref{eq:L-prime}.
These compositions are determined by
two jump superoperators:
$\hat{\mathcal{K}}^{(+)}_{B_{+}}$ and $\hat{\mathcal{K}}^{(-)}_{B_{-}}$,
and one has to find the matrices, $B_{+}$ and $B_{-}$, such that,
in the transformed Liouvillian
\begin{align}
  \label{eq:L-diag}
  \hat{\mathcal T}\hat{\mathcal{L}}\hat{\mathcal T}^{-1}=&
      -i\hat{\mathcal{N}}^{(-)}_{\Omega'}+2 \hat{\mathcal{K}}^{(0)}_{\Gamma_{0}'}
      \notag
  \\
  &
      +\Tr\Gamma\,\hat{\mathcal{I}}\equiv  \hat{\mathcal L}_d,\quad
      \Gamma_{0}'<0
\end{align}
both matrices associated with $\hat{\mathcal{K}}^{(+)}$ and
$\hat{\mathcal{K}}^{(-)}$ vanish.
An important point is that the operator of evolution
$\e^{\hat{\mathcal{L}}_d t}$ is required to be bounded from above at $t\ge 0$.
This places the above constraint $\Gamma_0'<0$ on
the relaxation matrix to ensure that
real parts of eigenvalues of $\hat{\mathcal{L}}_d$ are nonpositive.

From Eq.~\eqref{eq:Gamma_nu-prime}, 
the conditions $\Gamma_{\pm}'=0$
will require solving algebraic Riccati equations
that, in general, do not have closed-form solutions.
Theoretical considerations detailed in Appendix~\ref{app:Riccati}
show that, in our case, the Riccati equations can be treated
analytically. In what follows, we will utilize
the results of Appendix~\ref{app:Riccati} to derive
both the jump-eliminating transformations and the effective Hamiltonian.

We begin with the transformation of the form
\begin{align}
  \label{eq:T_pm-matr}
    \hat{\mathcal T}_{+-}(A_{+},A_{-})=e^{\hat{\mathcal{K}}^{(+)}_{A_{+}}}e^{\hat{\mathcal{K}}^{(-)}_{A_{-}}}.
\end{align}
This is the case where
the condition $\Gamma_{-}'=0$
leads to the Riccati equation~\eqref{eq:app-R1}
with $\nu=-1$
and its solution can be taken in
the simplest form
$A_{-}=I$ (see Eq.~\eqref{eq:app-Anu-I}).
Then the resulting effective non-Hermitian Hamiltonian
with $\Gamma_0'=-\Gamma$
is given by (see Eq.~\eqref{eq:app-Leff})
\begin{align}
  &
      \label{eq:LdHeff}
    \hat{\mathcal L}_d=
    -i(\overleftarrow{\hat{H}_{\mathrm{eff}}}-\overrightarrow{\hat{H}_{\mathrm{eff}}^{\dagger}})=
    \overleftarrow{\hat{L}_{\mathrm{eff}}}+\overrightarrow{\hat{L}_{\mathrm{eff}}^{\dagger}},
  \\
  &
    \label{eq:Heff}
  \hat{H}_{\mathrm{eff}}=\sum_{nm}(\Omega-i\Gamma)_{nm}\hat{a}^\dagger_n\hat{a}_m
  \notag
  \\
  &
  =
  \hat{J}_{H},\quad 
    H=\Omega-i\Gamma,
  \\
  &
    \label{eq:Leff}
     \hat{L}_{\mathrm{eff}}=-i\hat{H}_{\mathrm{eff}}=\hat{J}_{L},\quad 
    L=-i\Omega-\Gamma.
\end{align}

Since $\Gamma_{-}'=0$,
the jump-eliminating condition for the remaining recycling term
leads to a linear equation for the matrix $A_{+}$
of the form~\eqref{eq:Wnu-eq} 
\begin{align}
  \label{eq:Ap-eq}
  &
    \Gamma_{+}+\frac{i}{2}[\Omega',A_{+}]-
    \frac{1}{2}\{\Gamma_0',A_{+}\}
    \notag
  \\
  &
    =\Gamma_{+}-(L A_{+}+A_{+}\hcnj{L})/2=0.
\end{align}
According to Appendix~\ref{app:Riccati}
(see Eqs.~\eqref{eq:Wnu-t}--~\eqref{eq:Wnu-sol})
its solution reads
\begin{align}
  \label{eq:Ap-sol}
  A_{+}=-W_{+}=-2\int_0^{\infty}e^{L \tau}\Gamma_{+}e^{\hcnj{L}\tau}\dd\tau.
\end{align}
An important point is that the matrix $W_{+}$ is closely related to
the steady-state (equilibrium) density matrix
$\hat{\rho}_{\ind{ss}}$
($\hat{\mathcal{L}}\hat{\rho}_{\ind{ss}}=0$).
More precisely,
it can be shown~\cite{Kiselev:symmetry:2021}
that
the steady-state normally-ordered characteristic function
is Gaussian
$\chi_{\ind{ss}}(\bs{\alpha})=\exp\{-\sum_{m,n}W_{mn}^{(+)}\cnj{\alpha}_m\alpha_n\}$
and the corresponding steady-state statistical operator is given by
\begin{align}
  \label{eq:rho_ss}
  \hat{\rho}_{\ind{ss}}=\frac{\exp\{-\hat{J}_{V_{\ind{ss}}}\}}{
  \Tr\exp\{-\hat{J}_{V_{\ind{ss}}}\}
  },
  \quad
  e^{V_{\ind{ss}}}=(W_{+})^{-1}+I.
\end{align}

For the differently ordered transformation
\begin{align}
  \label{eq:T_mp-matr}
    \hat{\mathcal T}_{-+}(B_{-},B_{+})=e^{\hat{\mathcal{K}}^{(-)}_{B_{-}}}e^{\hat{\mathcal{K}}^{(+)}_{B_{+}}}
\end{align}
we cannot use Eq.~\eqref{eq:app-Anu} because, at $\nu=1$,
the relaxation matrix $\Gamma_0'$ (see Eq.~\eqref{eq:Omg-G0-prime-1})
is no longer negative definite.
In this case,
the properly defined effective Hamiltonian
can be obtained using
the solution given by Eq.~\eqref{eq:app-Anu-2}.
By using relation~\eqref{eq:app-Amnu},
similar to Eq.~\eqref{eq:Ap-sol},
the matrix $B_{+}$
\begin{align}
  \label{eq:Bm}
  B_{+}=(W_{-})^{-1}-I=-((W_{+})^{-1}+I)^{-1}=-e^{-V_{\ind{ss}}}
\end{align}
can be expressed in terms of $W_{+}$,
whereas equation for the matrix $B_{-}$
\begin{align}
  \label{eq:Bm-eq}
  &
    \Gamma_{-}+\frac{i}{2}[\Omega',B_{-}]+
    \frac{1}{2}\{\Gamma_0',B_{-}\}=0
\end{align}
can be treated using
the method developed in Appendix~\ref{app:Riccati}
for solving Eq.~\eqref{eq:Wnu-eq}.

We can now apply the above general results
to thermal bath Liouvillians with
the relaxation matrices given by Eq.~\eqref{eq:Gamma_pm}.
This is the special case of our primary concern
that was briefly discussed at the end of Appendix~\ref{app:Riccati}.
In this case, the matrices $W_{+}$ and $W_{-}$
are proportinal to the unity matrix:
$W_{+}=n_T I$ and $W_{-}=(n_T+1) I$.
As a result, a similar remark applies to the matrices
$A_{\pm}$ and $B_{\pm}$:
$A_{\pm}=\alpha_{\pm} I$ and $B_{\pm}=\beta_{\pm} I$.
So, the  differently ordered jump-eliminating
transformations can be written in the following simplified form:
\begin{align}
  \label{eq:Tpm}
  &
    \hat{\mathcal T}_{+-}(\alpha_{+},\alpha_{-})\equiv \hat{\mathcal
    T}_{+-}=e^{\alpha_{+}\hat{\mathcal{K}}^{(+)}_{I}}e^{\alpha_{-}\hat{\mathcal{K}}^{(-)}_{I}},
  \\
  &
    \alpha_{+}=-n_T,\quad \alpha_{-}=1,
  \\
  &
    \label{eq:Tmp}
        \hat{\mathcal T}_{-+}(\beta_{-},\beta_{+})\equiv \hat{\mathcal
    T}_{-+}=e^{\beta_{-}\hat{\mathcal{K}}^{(-)}_{I}}e^{\beta_{+}\hat{\mathcal{K}}^{(+)}_{I}},
  \\
  &
    \beta_{+}=-e^{-z_T},\quad
    \beta_{-}=n_T+1=\frac{1}{1-e^{-z_T}}\equiv Z
\end{align}
and
the resulting diagonalized Liouvillian~\eqref{eq:LdHeff} is given by
\begin{align}
  &
    \label{eq:Ld}
    \hat{\mathcal L}_{d}=
    \hat{\mathcal T}_{+-}(-n_T,1)\hat{\mathcal L}\hat{\mathcal T}_{+-}^{-1}(-n_T,1)
  \notag
    \\
    &
    =
      \hat{\mathcal T}_{-+}( Z,-e^{-z_T})\hat{\mathcal L}\hat{\mathcal
      T}_{-+}^{-1}( Z,-e^{-z_T})
      \notag
  \\
  &
        =
    \overleftarrow{\hat{L}_{\mathrm{eff}}}+\overrightarrow{\hat{L}_{\mathrm{eff}}^{\dagger}},
    \quad
    \hat{L}_{\mathrm{eff}}=-\hat{J}_{i\Omega+\Gamma}.
\end{align}


In contrast to
the Hamiltonian~\eqref{eq:Heff_0}
that
depends on temperature
and determines the so-called semiclassical limit
where the action of quantum jumps is neglected~\cite{Minganti:pra:2019},
the effective Hamiltonian~\eqref{eq:Heff}
is related to the diagonalized Liouvillian~\eqref{eq:Ld}
and is temperature independent
(it does not depend on $n_T$).
Interestingly, in the zero-temperature limit
with $n_T=0$, the difference between the Hamiltonians vanishes.

For $\hat{\mathcal{L}}^{\sharp}$, we have the relation
\begin{align}
  &
    \label{eq:Ld-sharp}
    \hat{\mathcal L}_{d}^{\sharp}=
    \hat{\mathcal T}_{-+}(n_T,-1)\hat{\mathcal L}^{\sharp}\hat{\mathcal T}_{-+}^{-1}(n_T,-1)
  \notag
    \\
    &
    =
      \hat{\mathcal T}_{+-}( -Z,e^{-z_T})\hat{\mathcal L}^{\sharp}\hat{\mathcal T}_{+-}^{-1}(
      -Z,e^{-z_T})
      \notag
  \\
  &
    =
    -i(\overrightarrow{\hat{H}_{\mathrm{eff}}}-\overleftarrow{\hat{H}_{\mathrm{eff}}^{\dagger}})=
    \overrightarrow{\hat{L}_{\mathrm{eff}}}+\overleftarrow{\hat{L}_{\mathrm{eff}}^{\dagger}}
\end{align}
that can be readily obtained from the conjugated versions of
Eqs.~\eqref{eq:Ld} and~\eqref{eq:LdHeff}
with the help of identities
\begin{align}
  &
  \label{eq:T-inv}
  \hat{\mathcal T}_{\pm\mp}^{-1}(\alpha_{\pm},\alpha_{\mp})=
    \hat{\mathcal T}_{\mp\pm}(-\alpha_{\mp},-\alpha_{\pm}),
  \\
  &
    \label{eq:T-cnj}
     \hat{\mathcal T}_{\pm\mp}^{\sharp}(\alpha_{\pm},\alpha_{\mp})=
    \hat{\mathcal T}_{\pm\mp}(\alpha_{\mp},\alpha_{\pm}).
\end{align}


\subsection{Spectrum, eigenoperators and superpropagators}
\label{subsec:spectrum}


Owing to Eq.~\eqref{eq:Ld},
the spectral problem for the Liouvillian~\eqref{eq:L-gen-form}
\begin{align}
  &
    \label{eq:L-lambda}
    \hat{\mathcal L}\hat{\rho}_{\lambda}=\lambda \hat{\rho}_{\lambda},
\quad \lambda\in\mathrm{spec}(\hat{\mathcal L})
\end{align}
is  directly related to the eigenvalue problem
for its diagonalized counterpart~\eqref{eq:LdHeff}
\begin{align}
  &
    \label{eq:Ld-lambda}
    \hat{\mathcal L}_{d}\,\hat{\rho}_{\lambda}^{(d)}=\lambda \hat{\rho}_{\lambda}^{(d)}. 
\end{align}
that can be easily solved provided the spectrum and eigenstates
of the effective Hamiltonian~\eqref{eq:Heff} are known.
More specifically,
given the eigenstates, $\ket{\mu}$,
and eigenvalues, $\mu$,
of $\hat{L}_{\ind{eff}}=-i\hat{H}_{\ind{eff}}$
with $\hat{L}_{\ind{eff}}\ket{\mu}=\mu\ket{\mu}$,
the relations
\begin{align}
  \label{eq:Leff-spec}
  \hat{\rho}_{\lambda}^{(d)}=\ket{\mu}\bra{\nu},
  \quad
  \lambda=\mu+\cnj{\nu}
\end{align}
where the asterisk denotes complex conjugation,
give the eigenmode $\hat{\rho}_{\lambda}^{(d)}$
and the corresponding eigenvalue.
In what follows, we will often use the operator
$\hat{L}_{\ind{eff}}=-i\hat{H}_{\ind{eff}}$
(see Eq.~\eqref{eq:Leff}) as a convenient substitute for $\hat{H}_{\ind{eff}}$.

We can now express the eigenoperator $\rho_{\lambda}$ in terms of
$\rho_{\lambda}^{(d)}$ using the superoperators
\begin{align}
  &
  \label{eq:Tpm-Tmp-inv}
    \hat{\mathcal T}_{+-}^{-1}\equiv\hat{\mathcal T}_{+-}^{-1}(-n_T,1)=\hat{\mathcal T}_{-+}(-1,n_T),
    \notag
  \\
  &
      \hat{\mathcal T}_{-+}^{-1}\equiv\hat{\mathcal T}_{-+}^{-1}( Z,-e^{-z_T})=\hat{\mathcal T}_{+-}(e^{-z_T},-Z)
\end{align}
that enter Eq.~\eqref{eq:Ld} as follows
\begin{align}
  &
  \label{eq:rho-lambda}
    \hat{\rho}_{\lambda}=\hat{\mathcal T}_{-+}^{-1}\hat{\rho}_{\lambda}^{(d)}
    =q_{\lambda}\hat{\mathcal T}_{+-}^{-1}\hat{\rho}_{\lambda}^{(d)},
\end{align}
where $q_{\lambda}$ is the coefficient of proportionality
takes into account the difference between images of
$\hat{\mathcal T}_{-+}^{-1}$ and $\hat{\mathcal T}_{+-}^{-1}$ that both belong to the one-dimensional
eigenspace.
Thus, the spectrum, $\mathrm{spec}(\hat{\mathcal L})$,
and the eigenoperators, $\hat{\rho}_{\lambda}$, are
governed by the spectral properties of the effective Hamiltonian~\eqref{eq:Heff}.

Similarly,  for eigenoperators of $\hat{\mathcal L}^{\sharp}$
and $\hat{\mathcal L}^{\sharp}_d$, we have
\begin{align}
  \label{eq:sigma-lambda1}
  &
    \hat{\mathcal
    L}^{\sharp}\hat{\sigma}_{\lambda}=\lambda\hat{\sigma}_{\lambda},
    \quad
    \hat{\mathcal
    L}_d^{\sharp}\hat{\sigma}_{\lambda}^{(d)}=\lambda\hat{\sigma}_{\lambda}^{(d)},
  \\
  &
     \label{eq:sigma-lambda2}
    \hat{\sigma}_{\lambda}=\hat{\mathcal T}_{+-}^{\sharp}\hat{\sigma}_{\lambda}^{(d)}.
\end{align}
Note that, when
the eigenmodes of the diagonalized Liouvillian and its conjugate
are normalized using
the biorthogonality condition
$\Tr\{\hat{\sigma}_{\lambda'}^{(d)}\hat{\rho}_{\lambda}^{(d)}\}=\delta_{\lambda'\lambda}$,
the biorthogonality relations
for $\hat{\rho}_{\lambda}$ and $\hat{\sigma}_{\lambda'}$
given by
\begin{align}
  \label{eq:q-lambda}
  &
    \Tr\{\hat{\sigma}_{\lambda'}
    \hat{\rho}_{\lambda}\}=
    \Tr\{
    \hat{\mathcal T}_{+-}^{\sharp}[\hat{\sigma}_{\lambda'}^{(d)}]
    \hat{\mathcal T}_{-+}^{-1}\hat{\rho}_{\lambda}^{(d)}
    \}
    \notag
  \\
  &
    =q_{\lambda}\Tr\{
    \hat{\mathcal T}_{+-}^{\sharp}[\hat{\sigma}_{\lambda}^{(d)}]
    \hat{\mathcal T}_{+-}^{-1}\hat{\rho}_{\lambda}^{(d)}
    \}=
    q_{\lambda}\delta_{\lambda'\lambda}
\end{align}
depend on the coefficient $q_{\lambda}$.


\subsubsection{Single-mode spectral problem}
\label{subsubsec:single}

In order to illustrate the basic steps of our method,
it is instructive
to consider the simplest single-mode case with $N=1$, where
$\mathcal{K}^{(\pm)}_{I}\equiv\mathcal{K}^{(\pm)}$ and
$\hat{L}_{\mathrm{eff}}=-(i\Omega+\Gamma)\hat{a}^\dagger\hat{a}$.

In this case,
the solution of the spectral problem
for the diagonalized Liouvillian
reads
\begin{subequations}
  \label{eq:1D-spec-d}
\begin{align}
  &
    \label{eq:1D-lambda}
    \lambda_{mn}=-i\Omega(m-n)-\Gamma(m+n),
  \\
  &
    \label{eq:1D-rho-d}
  \hat{\rho}_{\lambda}^{(d)}=\hat{\rho}_{mn}^{(d)}=\ket{m}\bra{n}\equiv\hat{\Pi}_{mn},
  \\
  &
    \label{eq:1D-sigma-d}
    \hat{\sigma}_{\lambda}^{(d)}=\hcnj{[\hat{\rho}_{mn}^{(d)}]}=
    \hat{\Pi}_{nm}=
    \hat{\sigma}_{nm}^{(d)}
\end{align}
\end{subequations}
whereas the eigenoperators
\begin{align}
  \label{eq:rho-mn-gen-1D}
    \hat{\rho}_{mn}=e^{e^{-z_T}\hat{\mathcal{K}}^{(+)}}e^{-Z\hat{\mathcal{K}}^{(-)}}\ket{m}\bra{n},
\end{align}
can be evaluated using the algebraic identity
\begin{align}
  &
  \label{eq:rho-mk-ident}
  e^{-Z\hat{\mathcal{K}}^{(-)}}\ket{m}\bra{n}=
  \sum_{k=0}^{\min(m,n)}(-Z)^k\sqrt{\binom{m}{k}\binom{n}{k}}
  \notag
  \\
  &
  \times\ket{m-k}\bra{n-k}.
\end{align}
The result
\begin{align}
  &
  \label{eq:rho-mk-1D}
  \hat{\rho}_{mn}=
  \sum_{k=0}^{\min(m,n)}\frac{(-Z)^k\sqrt{m!n!}}{k!(m-k)!(n-k)!}
    (\hat{a}^{\dagger})^{m-k}\hat{\rho}_0\hat{a}^{n-k}
\end{align}
can be conveniently expressed in terms of
the steady state eigenoperator
\begin{align}
  &
  \label{eq:rho_0-1D}
    \hat{\rho}_0=e^{e^{-z_T}\hat{\mathcal{K}}^{(+)}}\ket{0}\bra{0}=
    e^{-z_T\hat{a}^{\dagger}\hat{a}}=:e^{-Z^{-1}\hat{a}^{\dagger}\hat{a}}:,
\end{align}
where $:\ldots:$ stands for normal ordering, 
proportional to the density matrix of the thermal equilibrium state:
$\hat{\rho}_{\mathrm{eq}}=Z^{-1}\hat{\rho}_0$.
The expression for the eigenoperator,
$\hat{\sigma}_{\lambda}=\hat{\sigma}_{nm}$,
of the adjoint Liouvillian
$\hat{\mathcal L}^{\sharp}$
with the eigenvalue
$\lambda=\cnj{\lambda}_{nm}=\lambda_{mn}$
\begin{align}
  \label{eq:sigma-1D}
  &
    \hat{\sigma}_{nm}=\hat{\mathcal T}_{+-}^{\sharp}(-n_T,1)\hat{\Pi}_{nm}=
    \hat{\mathcal T}_{+-}(1,-n_T)\hat{\Pi}_{nm}
    \notag
  \\
  &
    =
    \sum_{k=0}^{\min(m,n)}\frac{(-n_T)^k\sqrt{m!n!}}{k!(m-k)!(n-k)!}
    (\hat{a}^{\dagger})^{n-k}\hat{a}^{m-k}  
\end{align}
can be derived with the help of Eq.~\eqref{eq:rho-mk-ident}
and the relation
\begin{align}
  &
  \label{eq:unity-1D}
    e^{\hat{\mathcal{K}}^{(+)}}\ket{0}\bra{0}=\sum_{n=0}^{\infty}\ket{n}\bra{n}=\hat{I}.
\end{align}

We can now briefly discuss
how the solution of the spectral problem
can be used to describe the evolution of
the density matrix in terms of eigenmodes.
To this end, in Appendix~\ref{app:biorth}, we have derived 
the biorthogonality relations for the eigenoperators~\eqref{eq:rho-mk-1D}
and~\eqref{eq:sigma-1D} (see Eq.~\eqref{eq:biorth-rel})
\begin{align}
    &
      \label{eq:qmn-1D}
      \mathrm{Tr}\{\hat{\sigma}_{n'm'}\hat{\rho}_{mn}\}=q_{mn}\delta_{nn'}\delta_{mm'},
      \quad q_{mn}=Z^{m+n+1}.
\end{align}
So, the eigenmode  expansion describing dynamics
of the density matrix takes the following form:
\begin{align}
    &
  \label{eq:evol-1D}
      \hat{\rho}(t)=\sum_{m,n=0}^{\infty} \frac{e^{\lambda_{mn} t}}{q_{mn}}\hat{\rho}_{mn}
      \mathrm{Tr}\{\hat{\sigma}_{nm}\hat\rho(0)\}.
\end{align}




\subsubsection{Multi-mode spectral problem}
\label{subsubsec:multi}

By contrast to the single-mode case,
solving the
spectral problem for the diagonalized
multimode Liouvillian~\eqref{eq:LdHeff} requires additional considerations
due to intermode interactions that complicate
the effective Hamiltonian~\eqref{eq:Heff} so that
the associated matrix $H$ is non-diagonal.
We shall apply the algebraic relations
\begin{align}
  &
  \label{eq:ident-Hpm-Vpm}
    e^{\overleftarrow{\hat{V}_{+}}+\overrightarrow{\hat{V}_{-}}}
  \{\overleftarrow{\hat{H}_{+}}+\overrightarrow{\hat{H}_{-}}\}
  e^{-\overleftarrow{\hat{V}_{+}}-\overrightarrow{\hat{V}_{-}}}=
  \overleftarrow{\hat{H}_{+}'}+\overrightarrow{\hat{H}_{-}'},
  \notag
  \\
  &
    \hat{H}_{\pm}'=e^{\pm\hat{V}_{\pm}}\hat{H}_{\pm}e^{\mp\hat{V}_{\pm}}
\end{align}
to the special case with
$\hat{H}_{+}=\hat{H}_{-}^{\dagger}=\hat{L}_{\mathrm{eff}}=\hat{J}_L$,
so as to find the operator
$\hat{V}_{+}=\hat{V}_{-}^{\dagger}\equiv \hat{V}=\hat{J}_V$
that transforms the Hamiltonian into the diagonal form
(see Eq.~\eqref{eq:Jordan-ident-1}):
\begin{align}
  &
  \label{eq:H-eff-d}
    e^{\hat{V}}\hat{L}_{\mathrm{eff}}e^{-\hat{V}}=
    e^{\hat{J}_V}\hat{J}_{L}e^{-\hat{J}_V}=
    \hat{J}_{L_d}\equiv\hat{L}_{\mathrm{eff}}^{(d)},
  \notag
  \\
  &
L_d=e^V L e^{-V}=\mathrm{diag}(L^{(d)}_1,\ldots,L^{(d)}_N).
\end{align}
The resulting diagonalized effective Hamiltonian is
\begin{align}
  \label{eq:H_eff_diag}
  \hat{L}_{\mathrm{eff}}^{(d)}=
  \sum_{k=1}^N L^{(d)}_k
  \hat{a}^\dagger_k\hat{a}_k=
  -\sum_{k=1}^N\Bigl(i\Omega^{(d)}_k+\Gamma^{(d)}_k\Bigr)
  \hat{a}^\dagger_k\hat{a}_k,
\end{align}
where $\Omega^{(d)}_k =-\Im L_k^{(d)}$
and
$\Gamma^{(d)}_k=-\Re L_k^{(d)}$.
Note that, for creation operators, the corresponding
mapping is given by Eq.~\eqref{eq:Jordan-ident-2}.

For the diagonalized Liouvillian superoperator
\begin{align}
  \label{eq:L-eff-d}
\hat{\mathcal{L}}_{\mathrm{eff}}^{(d)}=  
  \overleftarrow{\hat{L}_{\mathrm{eff}}^{(d)}} + \overrightarrow{\hcnj{[\hat{L}_{\mathrm{eff}}^{(d)}]}}
\end{align}
governed by the effective Hamiltonian~\eqref{eq:H_eff_diag},
eigenvalues and the corresponding eigenoperators of
are given by
\begin{align}
  &
    \label{eq:spec-Heff}
    \lambda_{\mathbf{m n}}=-\sum_{k=1}^N\Bigl\{i\Omega^{(d)}_k(m_k-n_k)
    \notag
  \\
  &
    +\Gamma^{(d)}_k (m_k+n_k)\Bigr\},
    \quad
    m_k,n_k\in\mathbb{N},
  \\
  &
    \label{eq:sigma_nm}
  \hat{\rho}_\lambda^{(d)}=
  \hat{\rho}_{\mathbf{mn}}^{(d)}
    =e^{-\hat{V}}\hat{\Pi}_{\mathbf{mn}}e^{-\hat{V}^{\dagger}},
\:
    \hat{\Pi}_{\mathbf{mn}}=
    \ket{\mathbf{m}}\bra{\mathbf{n}}
\end{align}
where
$\mathbf{m}\equiv(m_1,\ldots m_N)$
($\mathbf{n}\equiv(n_1,\ldots n_N)$)
is the multi-index
that labels
is the multimode Fock ket(bra)-state
$\ket{\mathbf{m}}=\ket{m_1}\otimes\ldots\otimes\ket{m_N}$
($\bra{\mathbf{n}}=\bra{n_1}\otimes\ldots\otimes\bra{n_N}$).

The matrix density eigenoperators given
by Eqs.~\eqref{eq:rho-mn-gen-1D}
and~\eqref{eq:sigma-1D}
that enter the expansion~\eqref{eq:evol-1D}
are generalized as follows
\begin{align}
  &
  \label{eq:rho-mn-gen}
  \hat{\rho}_{\mathbf{mn}}=e^{e^{-z_T}\hat{\mathcal{K}}^{(+)}_I}e^{-Z\hat{\mathcal{K}}^{(-)}_I}e^{-\hat{V}}\hat{\Pi}_{\mathbf{mn}}e^{-\hat{V}^{\dagger}},
  \\
  &
    \label{eq:sigma-nm-gen}
      \hat{\sigma}_{\mathbf{nm}}=e^{\hat{\mathcal{K}}^{(+)}_I}e^{-n_T\hat{\mathcal{K}}^{(-)}_I}e^{\hat{V}^{\dagger}}\hat{\Pi}_{\mathbf{nm}}e^{\hat{V}}
\end{align}
so that it is rather straightforward to write down
the multimode version of  Eq.~\eqref{eq:evol-1D}:
\begin{align}
    &
  \label{eq:evol-N}
      \hat{\rho}(t)=\sum_{\mathbf{m,n}}\frac{e^{\lambda_{\mathbf{mn}} t}}{q_{\mathbf{mn}}}\hat{\rho}_{\mathbf{mn}}
      \mathrm{Tr}\{\hat{\sigma}_{\mathbf{nm}}\hat\rho(0)\},
  \\
    &
      \label{eq:qmn-N}
      q_{\mathbf{mn}}=\mathrm{Tr}\{\hat{\sigma}_{\mathbf{nm}}\hat{\rho}_{\mathbf{mn}}\}.
\end{align}

Alternatively,
dynamics of the density matrix
can be described using the relations
\begin{align}
  &
    \label{eq:evol-Pimn}
    e^{\hat{\mathcal L}_d t}\ket{\mathbf{m}}\bra{\mathbf{n}}=
  e^{\hat{L}_{\mathrm{eff}}t}\ket{\mathbf{m}}\bra{\mathbf{n}}e^{\hat{L}_{\mathrm{eff}}^{\dagger}t}
  \notag
  \\
  &
  =\prod_{i=1}^N\frac{(\hat{b}_i^{\dagger})^{m_i}}{\sqrt{m_i!}}
    \ket{\vc{0}}\bra{\vc{0}}\prod_{i=1}^N\frac{\hat{b}_i^{n_i}}{\sqrt{n_i!}},
  \\
  &
    \label{eq:bc-vs-ac}
    \hat{b}_i^{\dagger}=e^{\hat{L}_{\mathrm{eff}}t}\hat{a}_i^{\dagger}e^{-\hat{L}_{\mathrm{eff}}t}=
    \sum_{j=1}^NP_{ji}(t)\hat{a}_j^{\dagger},
  \\
  &
    \label{eq:b-vs-a}
    \hat{b}_i=\sum_{j=1}^NP_{ji}^{*}(t)\hat{a}_j,\quad
    P(t)=e^{Lt}=e^{-(i\Omega+\Gamma/2)t}
\end{align}
in terms of the operators
\begin{align}
  &
  \label{eq:rho-mn-1}
    \hat{\rho}_{\mathbf{mn}}^{(1)}=e^{e^{-z_T}\hat{\mathcal{K}}^{(+)}_I}e^{-Z\hat{\mathcal{K}}^{(-)}_I}\hat{\Pi}_{\mathbf{mn}}
    =\prod_{i=1}^N\hat{\rho}_{m_i n_i},
  \\
  &
    \label{eq:sigma-nm-1}
    \hat{\sigma}_{\mathbf{nm}}^{(1)}=e^{\hat{\mathcal{K}}^{(+)}_I}e^{-n_T\hat{\mathcal{K}}^{(-)}_I}\hat{\Pi}_{\mathbf{nm}}
    =\prod_{i=1}^N\hat{\sigma}_{n_i m_i},
\end{align}
where $\hat{\rho}_{m_i n_i}$
and $\hat{\sigma}_{n_i m_i}$
are given by Eq.~\eqref{eq:rho-mk-1D} and Eq.~\eqref{eq:sigma-1D},
respectively,
and the matrix elements
\begin{align}
  \label{eq:U-mmprime}
U_{\mathbf{m'm}}(t)=\bra{\mathbf{m'}}e^{\hat{L}_{\mathrm{eff}}t}\ket{\mathbf{m}}
\end{align}
that can be evaluated using Eq.~\eqref{eq:bc-vs-ac}.

So, the dynamics of our system
is governed by
the action of the superpropagator
$e^{\hat{\mathcal L}t}$ on
the statistical operator
at the initial instant of time,
expressed in terms of
the products of single-mode operators
as follows:
\begin{align}
    &
  \label{eq:evol-N-1}
      \hat{\rho}(t)=e^{\hat{\mathcal L}t}\hat{\rho}(0)
      =\sum_{\mathbf{m,n,m',n'}}
      \frac{U_{\mathbf{m'm}}(t)U_{\mathbf{n'n}}^{*}(t)}{q_{\mathbf{m'n'}}^{(1)}}
      \notag
  \\
    &
      \times
      \hat{\rho}_{\mathbf{m'n'}}^{(1)}
      \mathrm{Tr}\{\hat{\sigma}_{\mathbf{nm}}^{(1)}\hat\rho(0)\},
  \\
    &
      \label{eq:qmn-N-1}
      q_{\mathbf{mn}}^{(1)}=\prod_{i=1}^Nq_{m_i n_i},
\end{align}
where
$q_{m_in_i}=\mathrm{Tr}\{\hat{\sigma}_{n_im_i}\hat{\rho}_{m_in_i}\}=Z^{n_i+m_i+1}$.

An important point to be emphasized is that,
in contrast to Eqs.~\eqref{eq:H_eff_diag}--\eqref{eq:qmn-N},
above formulas
(Eqs.~\eqref{eq:evol-Pimn}--\eqref{eq:qmn-N-1})
remain applicable
even when the matrices $H$ and $L$
for the effective Hamiltonian $\hat{H}_{\ind{eff}}$
and $\hat{L}_{\ind{eff}}$ are nondiagonalizable.
The latter is the condition for an exceptional point (EP) to occur
(validity of the spectrum~\eqref{eq:spec-Heff} rules out exceptional points).
At EPs,  the matrix exponential $e^{Lt}=e^{-iHt}$ describing the temporal evolution of the
creation and annihilation operators in
Eqs.~\eqref{eq:bc-vs-ac} and~\eqref{eq:b-vs-a}
are well defined and can be used to explore
the effects of critical slowdown associated with EPs.

\section{Applications: Exceptional points and speed of evolution}
\label{sec:applications}

The results of the previous section
demonstrate that our method
can be used as an efficient tool to
perform a comprehensive analysis of the
spectral properties of
the multi-mode bosonic Liouvillians~\eqref{eq:L-rho}.
As discussed in Sec.~\ref{sec:intro},
such analysis will be of crucial importance for understanding
physical phenomena near exceptional points.
More generally, the eigenspectrum and the eigenoperators
are intimately related to
the intermode couplings-induced regimes of the Lindblad dynamics.
One of the natural ways to deal with these regimes
is to study how the inter-mode interactions
influence the speed of evolution
underlying the concept of the quantum speed
limit (QSL) time.

The latter is
defined as the minimal time needed for a system to perform a transition
between predetermined initial and final (target) states.
Quantum mechanics is known to dictate QSL bounds for the minimal evolution time,
which is inversely related to the evolution speed.
For the case of unitary evolution, the well-known
Mandelstam-Tamm~\cite{Tamm:jphys:1945,Anandan:prl:1990} and
Margolus-Levitin~\cite{Uffink:ajp:1993,Margolus:physd:1998,Levitin:prl:2009} inequalities
provide general limits on the speed of dynamical evolution
and set a dynamical intrinsic time scale.
These inequalities have been extended to
the case of nonunitary dynamics of open quantum systems
using the geometric approach where
quantum speed limits are derived as upper bounds on the rate of
change of geometric measures of distinguishability
such as the relative purity~\cite{Campo:prl:2013},
the Bures angle~\cite{Deffner:prl:2013},
the quantum Fisher information~\cite{Taddei:prl:2013}  
(see also unifying considerations
presented in Refs.~\cite{Pires:prx:2016,Lan:njp:2022}).

It is rather straightforward to illustrate the basic steps of the
geometric approach assuming that the initial state is pure,
$\hat{\rho}(0)=\ket{\psi_0}\bra{\psi_0}$,
and the measure of distinguishability is determined by
the Uhlmann fidelity.
In this case, we can use 
the simplified expression for the fidelity
\begin{align}
  \label{eq:fidelity}
  F(\hat{\rho}(0),\hat{\rho}(t))=\bra{\psi_0}\hat{\rho}(t)\ket{\psi_0}\equiv F(t)
\end{align}
and, with the help of the Cauchy-Schwarz inequality,
show that the rate of change of the fidelity~\eqref{eq:fidelity}
\begin{align}
  &
  \label{eq:ineq-F}
  \Bigl|\frac{\partial F(t)}{\partial t}\Bigr|
  \le v(t)
 \end{align}
 is bounded from above  by the speed of evolution
 defined using
the Hilbert-Schmidt metric~\cite{brody2019evolution,Marian:pra:1:2021,Kiselev:entropy:2022}:
 \begin{align}
  &
      \label{eq:speed-def}
    v(t)=
    \sqrt{\Tr\Bigl[\frac{\partial \hat{\rho}(t)}{\partial t}\Bigr]^2}
    \equiv
    \Bigl\|\frac{\partial \hat{\rho}(t)}{\partial t}\Bigr\|_2.
 \end{align}
From Eq.~\eqref{eq:ineq-F}, we have the QSL inequality
 \begin{align}
   &
      \label{eq:QSL-F}
  t\ge t_F(t)=\frac{1-F(t)}{\avr{v}_t},
\quad    \avr{v}_t\equiv\frac{1}{t}\int_0^t v(\tau)\dd\tau
 \end{align}
for the fidelity-based time-dependent QSL time $t_F$.
 Note that similar reasoning is directly applicable to the case where
 the geometric measure of distinguishability
 is given by
 the Hilbert-Schmidt distance
 $\|\hat{\rho}(t)-\hat{\rho}(0)\|_2$ between the initial and 
 evolved states.

The  literature on quantum speed limits is quite voluminous,
and more details on QSLs and their applications in the fields such as
quantum metrology and optimal control theory can be found
in, e.g., Ref.~\cite{Deffner:jpa:2017}.
The study of QSLs is one of the hot areas in current research:
the limits for time-resolved sensing of dynamical signals using qubit
probes are found be closely related to the QSL~\cite{Herb:prl:2024};
bounds on the speed limit of quantum
evolution by unitary operators in arbitrary dimensions are established
in~\cite{Farmanian:pra:2024};
an attainable quantum speed limit based
on the variation of quantum coherence
using the skew information as the coherence measure
was obtained in~\cite{Mai:pra:2024};
the experimental verification to the
QSLs in various scenarios with a superconducting circuit
is presented in~\cite{Wu:pra:2024}.


For two-level open quantum systems that interact with
Markovian and non-Markovian environments,
the problem of evolution speed has been the subject of intense
studies~\cite{Cianciaruso:pra:2017,brody2019evolution,Haseli:ijtp:2020,Nie:ijtp:2021,Lan:njp:2022}.  A
systematic analysis of the most common quantum speed limit (QSL) bounds in the damped Jaynes-Cummings model was performed
in~\cite{Mirkin:pra:2016}. The problem of quantum metrology in the context of non-Markovian quantum
evolution of a two-qubit system was explored
in~\cite{Mirkin:pra:2020}.
Geometric quantum speed limits of a qubit subject to decoherence in an ensemble of
chloroform molecules were recently studied in a
nuclear magnetic resonance experiment~\cite{Pires:commphys:2024}.

Compared to the case of finite-dimensional quantum systems,
the evolution speed of
bosonic multi-mode quantum systems representing a family of
open continuous-variable systems
has received much less attention~\cite{Marian:pra:1:2021,Kiselev:entropy:2022,Wu:pra:2022}.
So, in this section, we show how the algebraic tools
developed in the previous sections
can be efficiently utilized to explore the problem
of the speed of evolution of
quantum polarization states
represented by bimodal bosonic systems~\cite{Goldberg:advopt:2021}.

Lindblad dynamics of such systems can be regarded as a model
describing propagation of quantized polarization modes in an optical
fiber~\cite{gaidash2021quantum}.
In this model,
the intermode coherent (incoherent) interactions
originate from the effects of
the birefringence (the dichroism or polarization-dependent losses).
In Sec.~\ref{subsec:two-mode}, we shall use
the frequency and relaxation vectors
to encode channel anisotropy parameters
characterizing the effective Hamiltonian.
In the parameter space determined by these vectors,
we deduce the conditions that govern the geometry of the exceptional points.
In Sec.~\ref{subsec:evolspeed},
by assuming that the mean number of photons, $n_T$,
is small,
approximate formulas for the Liouvillian and the evolution superpropagator
describing the low-temperature regime
in the first-order (linear) approximation
are employed to perform numerical analysis of
the evolution speed of
the one-photon polarization states known as
the polarization qubit.

\subsection{Exceptional points of two-mode systems}
\label{subsec:two-mode}

The effective Hamiltonian~\eqref{eq:Heff} of a two-mode system
is determined by the associated $2\times 2$ matrix, $H$.
This matrix is expressed in terms of the frequency and relaxation matrices,
$\Omega$ and $\Gamma$, that
can be written as linear combinations  of the Pauli matrices
\begin{align}
  &
  \label{eq:Omega-Gamma}
 H= {\Omega}-i\Gamma=
  \sum_{k=0}^3(\omega_k-i\gamma_k){\sigma}_k=
  (\omega_0-i\gamma_0){\sigma}_0
  \notag
  \\
  &
    +(\bs{\omega}-i\bs{\gamma},\bs{\sigma}),
    \quad
    {\sigma}_0\equiv \diag(1,1),
\end{align}
where
$\bs{\sigma}\equiv({\sigma}_1,{\sigma}_2,{\sigma}_3)$
is the vector of the Pauli matrices given by
\begin{align}
  &
    \label{eq:Pauli}
    {\sigma}_1=\begin{pmatrix}
    0 & 1 \\
    1 & 0
    \end{pmatrix}, \ 
    {\sigma}_2=\begin{pmatrix}
    0 & -i \\
    i & 0
    \end{pmatrix}, \ 
    {\sigma}_3=\begin{pmatrix}
    1 & 0 \\
    0 & -1
    \end{pmatrix},
\end{align}
$\bs{\omega}\equiv(\omega_1,\omega_2,\omega_3)$ and
$\bs{\gamma}\equiv(\gamma_1,\gamma_2,\gamma_3)$
is the frequency and the relaxation rate vector, respectively.
These vectors govern
the regime of dissipative dynamics
and
can be conveniently parameterized using the angular representation~\cite{gaidash2021quantum}:
\begin{align}
  &
  \label{eq:gamma-omega}
  \bs{\gamma}=\gamma(\sin\theta_{\Gamma}\cos\phi_{\Gamma},\sin\theta_{\Gamma}\sin\phi_{\Gamma},
  \cos\theta_{\Gamma})=\gamma\vc{n}_{\Gamma},
  \notag
  \\
  &
  \bs{\omega}=\omega(\sin\theta_{\Omega}\cos\phi_{\Omega},\sin\theta_{\Omega}\sin\phi_{\Omega},
    \cos\theta_{\Omega})
    =\omega\vc{n}_{\Omega}.
\end{align}

It is rather straightforward to obtain
the matrix exponential for ${P}(t)$
(see Eq.~\eqref{eq:b-vs-a})
in the following explicit form:
\begin{align}
  &
  \label{eq:P-2}
  {P}(t)=\ee^{-iH t}=
  \ee^{-(i\omega_0+\gamma_0)t}
  \Bigl\{\cosh(q t){\sigma}_0
  \notag
  \\
  &
  -
  \frac{\sinh(q t)}{q}(\bs{\gamma}+i\bs{\omega},\bs{\sigma})
  \Bigr\},
\end{align}
where
\begin{align}
  \label{eq:omega-gamma}
  &
  q=
\sqrt{(\bs{\gamma}+i\bs{\omega},\bs{\gamma}+i\bs{\omega})}=
  \sqrt{\sum_{k=1}^3(\gamma_k+i\omega_k)^2}.
\end{align}

From Eq.~\eqref{eq:P-2} it immediately follows
that exceptional points require
the rate parameter $q$ given by Eq.~\eqref{eq:omega-gamma}
to vanish, $q=0$.
Geometrically, this EP condition occurs when
the vectors $ \bs{\gamma}$ and $ \bs{\omega}$
are mutually orthogonal, $ \bs{\gamma}\perp\bs{\omega}$,
and are identical in length, $ |\bs{\gamma}|=|\bs{\omega}|$:
\begin{align}
  \label{eq:EP-cond}
  (\vc{n}_{\Gamma},\vc{n}_{\Omega})=0,
  \quad \gamma=\omega.
\end{align}
At EP,
the expression for the matrix exponential~\eqref{eq:P-2}
\begin{align}
  &
  \label{eq:P-EP}
  {P}(t)|_{q=0}=
  \ee^{-(i\omega_0+\gamma_0)t}
  \Bigl\{{\sigma}_0
  -
  \gamma t (\vc{n}_{\Gamma}+i\vc{n}_{\Omega},\bs{\sigma})
  \Bigr\}
\end{align}
shows that a rapidly varying combination of exponential functions
appear to be
replaced with a function linearly dependent on time.

\subsection{Speed of evolution of polarization qubit}
\label{subsec:evolspeed}

\begin{figure*}[!htp]
  \centering
  \begin{subfigure}{.3\textwidth}
    \centering
    \includegraphics[width=1\linewidth]{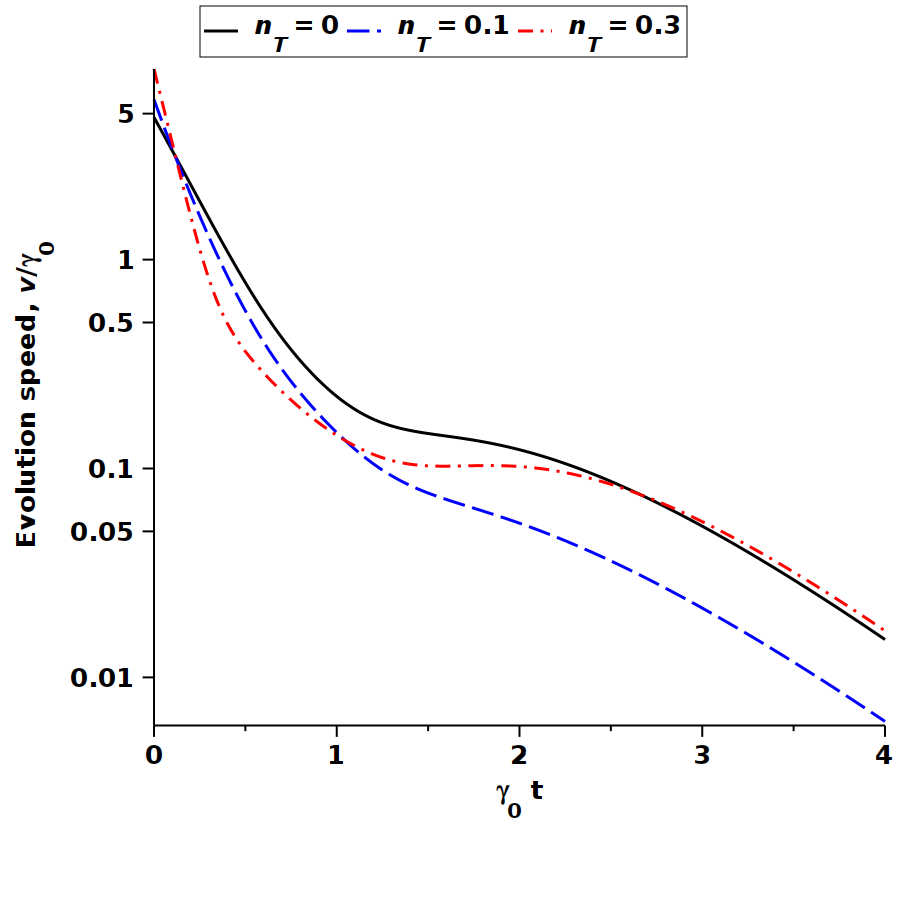}
    \caption{$\theta_{\Gamma}=\pi/2$}
    \label{subfig:1a}
  \end{subfigure}
  \begin{subfigure}{.3\textwidth}
    \centering
    \includegraphics[width=1\linewidth]{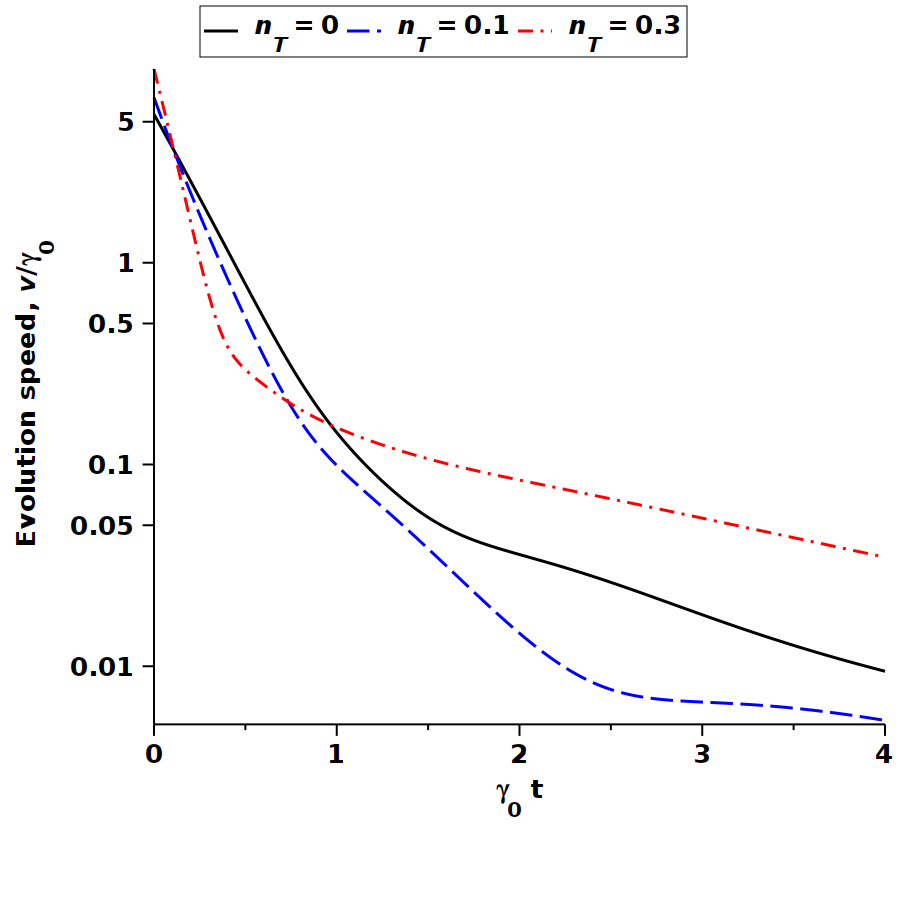}
    \caption{$\theta_{\Gamma}=\pi/4$}
    \label{subfig:1b}
  \end{subfigure}
   \begin{subfigure}{.3\textwidth}
    \centering
    \includegraphics[width=1\linewidth]{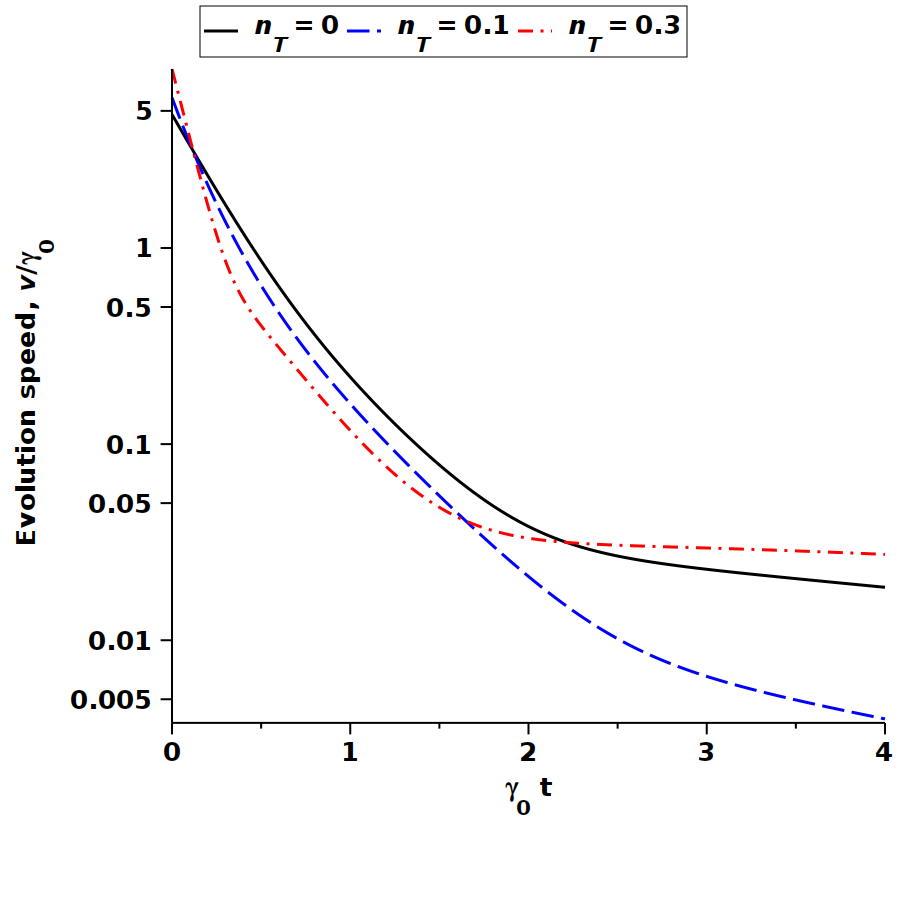}
    \caption{$\theta_{\Gamma}=0$}
    \label{subfig:1c}
  \end{subfigure}
\caption{
  Speed of evolution
  computed from Eq.~\eqref{eq:v2-nt}
  for the initial state~\eqref{eq:rho_0-0}
  with $\theta=\pi/4$ and $\phi=0$
  as a function of the dimensionless time parameter, $\gamma_0 t$,
  at the frequency and relaxation vectors:
  $\bs{\omega}=0.9\gamma_0(0,0,1)$ and
  $\bs{\gamma}=0.9\gamma_0(\sin\theta_{\Gamma},0,
  \cos\theta_{\Gamma})$.
  Three cases are shown:
  (a)~$\theta_{\Gamma}=\pi/2$, (b)~$\theta_{\Gamma}=\pi/4$
  and (c) $\theta_{\Gamma}=0$.
  Solid, dashed and dotted-dashed lines are evaluated
  at $n_T=0$, $n_T=0.1$ and $n_T=0.3$, respectively.
} 
\label{fig:tht_G}
\end{figure*}

In this section,
the dynamical behavior of the speed of evolution~\eqref{eq:speed-def}
which is governed by both the spectrum of the Liouvillian and
the initial state $\hat{\rho}(0)=\ket{\psi_0}\bra{\psi_0}$
will be our primary concern.
Since, for optical frequencies,
the mean number of thermal photons, $n_T$, is typically very small,
we concentrate on 
the low-temperature regime of the Lindblad dynamics. 

As was mentioned in
Sec.~\ref{subsec:Ldiag}, 
in the zero-temperature limit
with $n_T=0$,
the effective Hamiltonians
$H_{\ind{eff}}^{(0)}$ (see Eq.~\eqref{eq:Heff_0})
and
$H_{\ind{eff}}$
(see Eq.~\eqref{eq:Heff})
coincide.
At $n_T=0$, dynamics is governed by
the zero-temperature Liouvillian: $\hat{\mathcal L}_0=\hat{\mathcal L}|_{n_T=0}$.
Since $\hat{\mathcal T}_{-+}^{-1}|_{n_T=0}=\hat{\mathcal T}_{+-}(0,-1)$,
from Eq.~\eqref{eq:Ld}, we obtain the expression
for $\hat{\mathcal L}_0$ in the following form:
\begin{align}
  &
    \label{lo}
  \hat{\mathcal L}_0=e^{-\hat{\mathcal{K}}^{(-)}_{I}} \hat{\mathcal L}_d e^{\hat{\mathcal{K}}^{(-)}_{I}}.
\end{align}
According to Eq.~\eqref{eq:evol-Pimn},
the diagonalized Liouvillian
keeps the total number of photons
of the Fock states intact.
From the relations~\eqref{eq:rho-mk-ident},
it additionally follows
that the superoperator
$e^{-\hat{\mathcal{K}}^{(-)}_{I}}$ acting on the Fock states
cannot produce the Fock states with
a total number of photons larger than their initial values.
This suggests that the finite-dimensional
Fock subspace that contains the initial quantum state
remains invariant under the action of
the zero-temperature Liouvillian, $\hat{\mathcal L}_0$. 

 In the limiting case of
 zero-temperature environment with $n_T=0$,
 the squared evolution speed is given by
\begin{align}
  &
    \label{eq:v0-2}
    v_0^2(t)=
    \text{Tr}\Bigl\{(\hat{v}_0(t))^2\Bigr\},
    \quad
    \hat{v}_0\equiv \frac{\partial\hat{\rho}_0}{\partial t},
\end{align}
where
\begin{align}
  \label{eq:rho0-v0}
  &
    \hat{\rho}_0(t)=e^{\hat{\mathcal L}_0 t}\hat{\rho}(0)=
    ( e^{-\hat{\mathcal{K}}^{(-)}_{I}}e^{\hat{\mathcal L}_d t}
    e^{\hat{\mathcal{K}}^{(-)}_{I}})\hat{\rho}(0).    
\end{align}
When $\hat{\rho}(0)$ is the statistical operator
with the finite-dimensional support in the Fock basis,
the superoperator
$e^{-\hat{\mathcal{K}}^{(-)}_{I}}e^{\hat{\mathcal L}_{d}t} e^{\hat{\mathcal{K}}^{(-)}_{I}}$
keeps the support dimension finite and
computing of
the squared evolution speed does not require using sophisticated tools.

We can now use the approximate formula
\begin{align}
  &
  e^{\hat{\mathcal L}t}\approx\Big(\hat{\mathcal I}+n_T\hat{U}_1(t)\Big)e^{\hat{\mathcal L}_0 t},
    \label{linearres}
  \\
  &
    \label{eq:U1-out}
    \hat{U}_1(t)=
  \Delta\hat{\mathcal L}_{Q(t)}
    =\hat{\mathcal{K}}_{Q(t)}^{(+)}-2\hat{\mathcal{K}}_{Q(t)}^{(0)}+\hat{\mathcal{K}}_{Q(t)}^{(-)},
  \\
  &
    \label{eq:W}
  Q(t)=I-e^{Lt}e^{L^{\dagger}t}=I-P(t)P^{\dagger}(t)  
\end{align}
to study the effect of thermal photons on the evolution speed in
the first-order (linear) approximation.
Details on the calculation of this operator can be found
in Appendix~\ref{appder}.
The approximated superpropagator
generalizes the similar result
previously reported in~\cite{gaidash2020dissipative} for the single-mode model to the case of
multimode systems.
In addition, it is rather straightforward to derive higher-order corrections
using algebraic relations and transformations
described in the Appendix~\ref{appder}.

According to Eq.~\eqref{eq:evol-1},
in this approximation,
the density matrix can be written in the form
\begin{align}
  \label{eq:rho1-1}
  \hat{\rho}(t)\approx \hat{\rho}_0(t)+n_T\hat{\rho}_1(t),
  \quad \hat{\rho}_1(t)=\hat{U}_1(t)\rho_0(t),
\end{align}
where
the superoperator $\hat{U}_1(t)$ given by Eq.~\eqref{eq:U1-out}
acts on the zero-temperature density matrix, $\hat{\rho}_0$,
producing the first order correction, $\hat{\rho}_1$.
The resulting expression for the squared speed of evolution reads
\begin{align}
  &
    \label{eq:v2-nt}
    v^2(t)\approx \Tr((\hat{v}_0+n_T\hat{v}_1)^2),
    \quad
    \hat{v}_1\equiv \frac{\partial\hat{\rho}_1}{\partial t}.
\end{align}
  
The approximate relations~\eqref{eq:v0-2}
and~\eqref{eq:v2-nt} can now be applied to the special case of the two-mode system
introduced in Sec.~\ref{subsec:two-mode}.
To be specific,
it will be assumed
that the system represents the quantum polarization states,
where
$\hat{a}_1\equiv\hat{a}_H$ and $\hat{a}_2\equiv\hat{a}_V$
are the annihilation operators for horizontally and vertically polarized
photon modes, respectively,
and the initial state is
a pure single-photon state
taken in the form:
\begin{align}
  \label{eq:pol-qubit}
  &
    \hat{\rho}(0)=\ket{\psi_0}\bra{\psi_0},
    \notag
  \\
  &
  \ket{\psi_0}=\cos\frac{\theta}{2}\ket{1_H,0_V}+ e^{i\phi} \sin\frac{\theta}{2} \ket{0_H,1_V}
\end{align}
where $\ket{n_H,m_V}\equiv\ket{n_H}\otimes\ket{m_V}$.
This is the state of
polarization qubit which, at $\phi=0$, is linearly polarized
along the unit vector,
$(\cos\theta,\sin\theta,0)$,
specified by the polarization azimuth $\theta$.
The initial density matrix can be rewritten
using the superoperators
$\hat{\mathcal{K}}^{(+)}_{mn}$ as follows
\begin{align}
  \label{eq:rho_0-0}
  \hat{\rho}(0)=
  \hat{\mathcal{K}}^{(+)}_{R_0}
  \ket{\vc{0}}\bra{\vc{0}},
  \quad
  R_0=\frac{1}{2}\left\{\sigma_0+(\vc{n},\bs{\sigma})\right\}
\end{align}
where $\vc{n}=(\sin\theta\cos\phi,\sin\theta\sin\phi,\cos\theta)$ and
$\ket{\vc{0}}\equiv\ket{0_H,0_V}$ is the vacuum state.

\begin{figure*}[!htp]
  \centering
    \begin{subfigure}{.45\textwidth}
    \centering
    \includegraphics[width=1\linewidth]{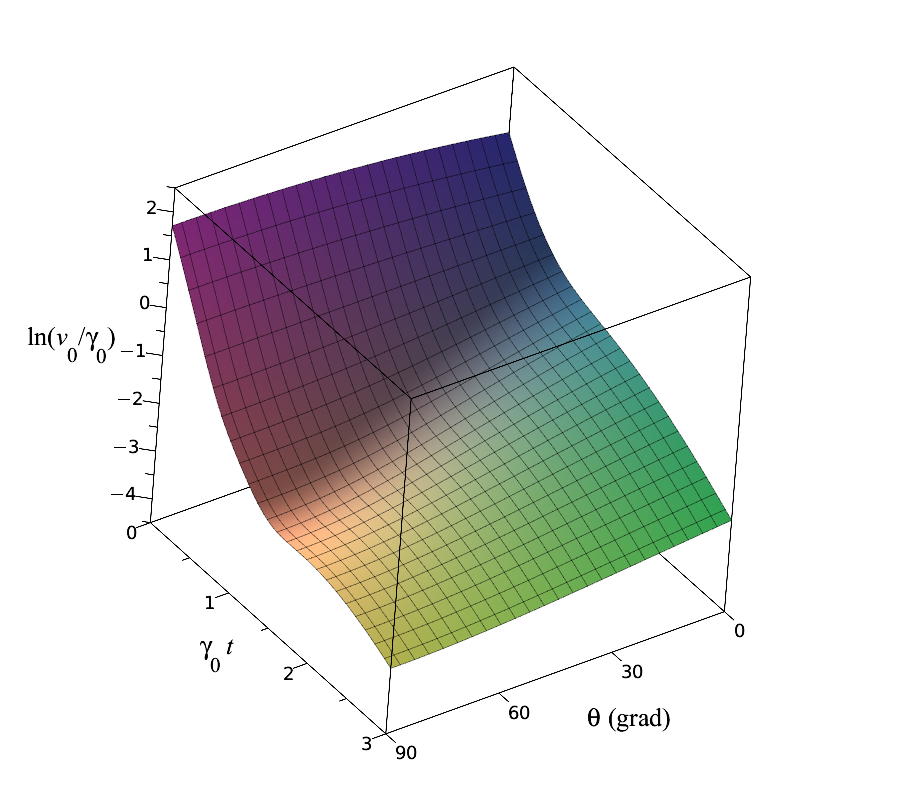}
    \caption{$n_T=0,\:\theta_{\Gamma}=\pi/2$}
    \label{subfig:2a}
  \end{subfigure}
  \begin{subfigure}{.45\textwidth}
    \centering
    \includegraphics[width=1\linewidth]{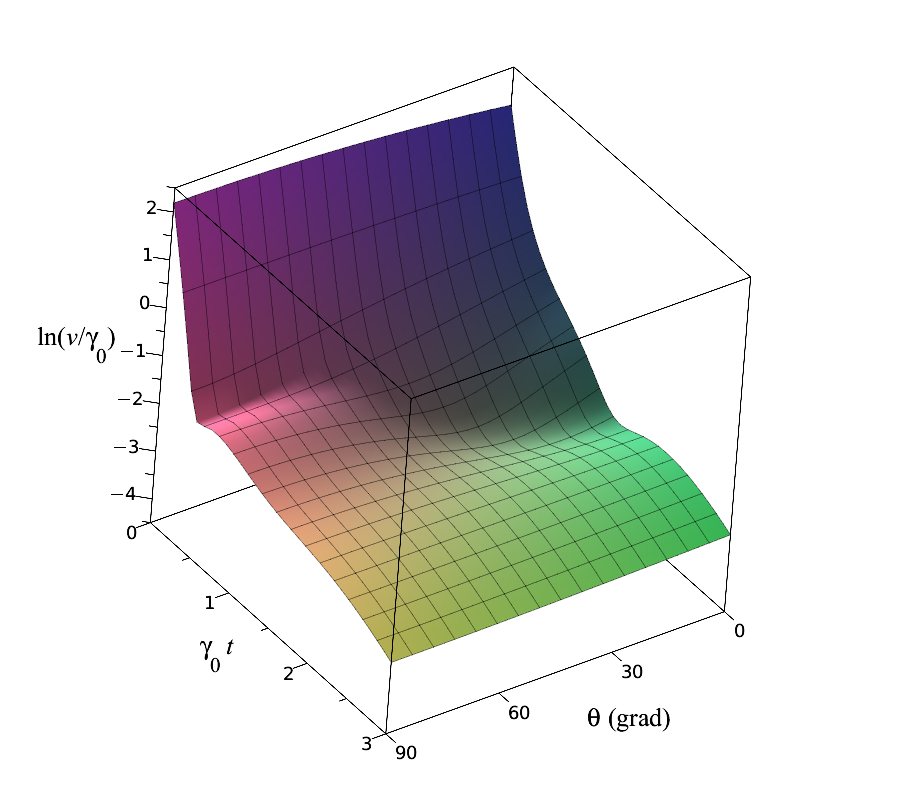}
    \caption{$n_T=0.3,\:\theta_{\Gamma}=\pi/2$}
    \label{subfig:2b}   
  \end{subfigure}
  \begin{subfigure}{.45\textwidth}
    \centering
    \includegraphics[width=1\linewidth]{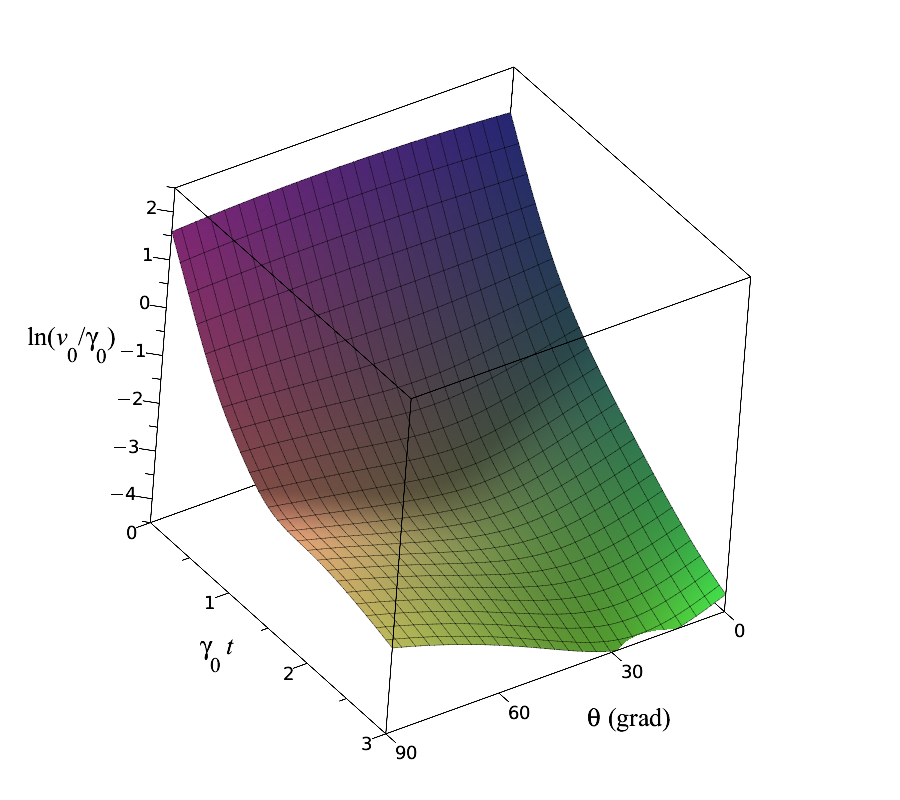}
    \caption{$n_T=0,\:\theta_{\Gamma}=\pi/4$}
    \label{subfig:2c}
  \end{subfigure}
  \begin{subfigure}{.45\textwidth}
    \centering
    \includegraphics[width=1\linewidth]{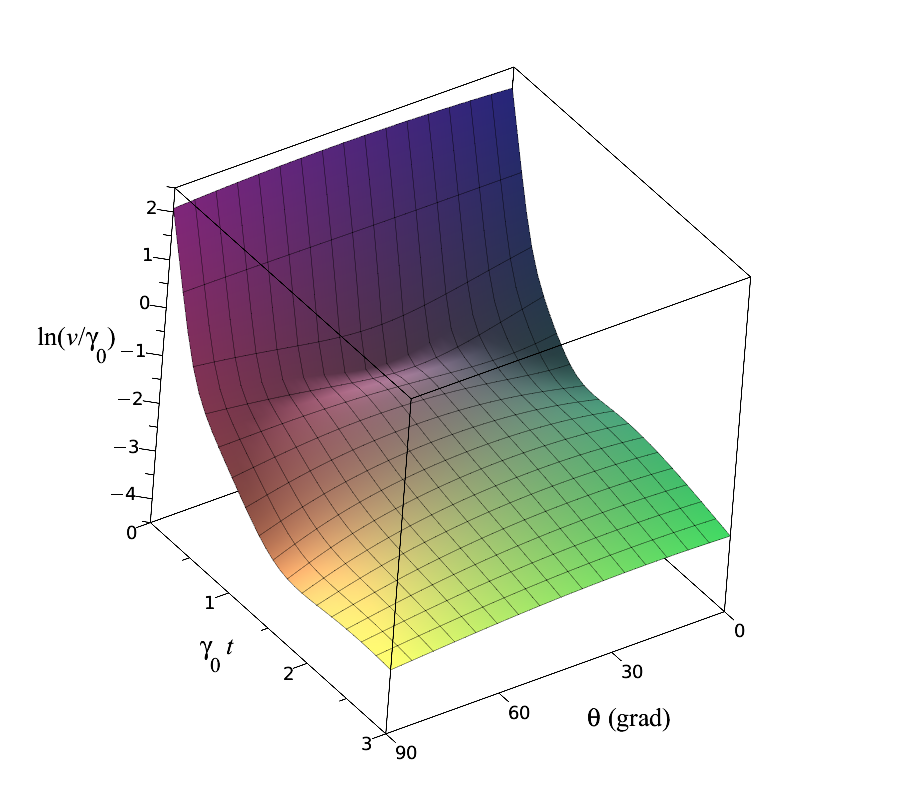}
    \caption{$n_T=0.3,\:\theta_{\Gamma}=\pi/4$}
    \label{subfig:2d}
  \end{subfigure}
    \begin{subfigure}{.45\textwidth}
    \centering
    \includegraphics[width=1\linewidth]{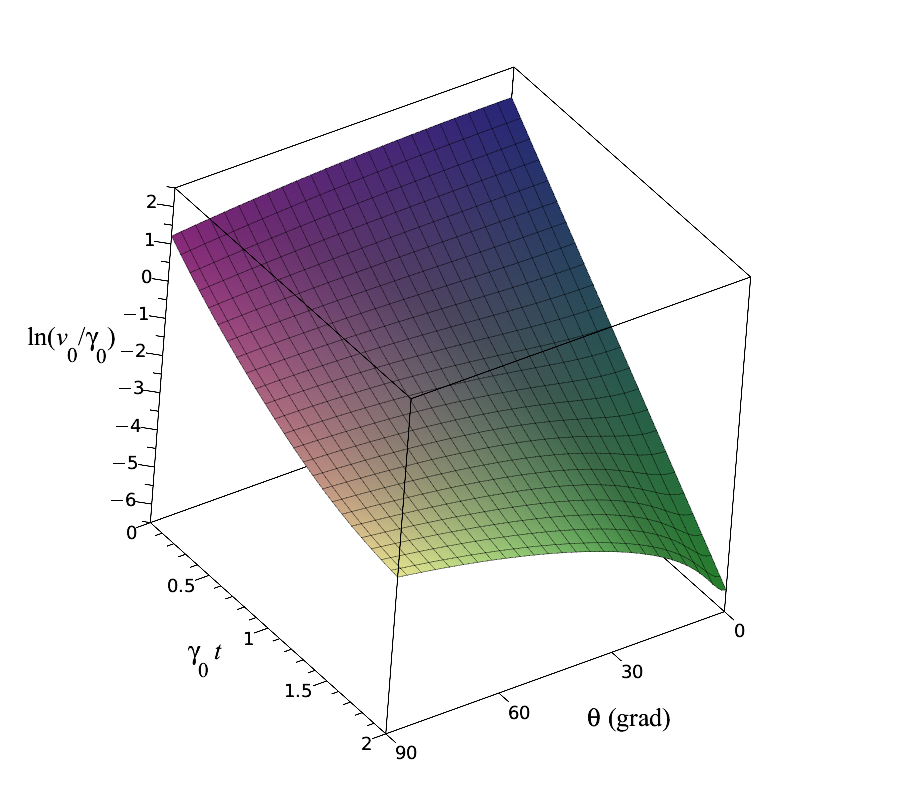}
    \caption{$n_T=0,\:\theta_{\Gamma}=0$}
    \label{subfig:2e}
  \end{subfigure}
  \begin{subfigure}{.45\textwidth}
    \centering
    \includegraphics[width=1\linewidth]{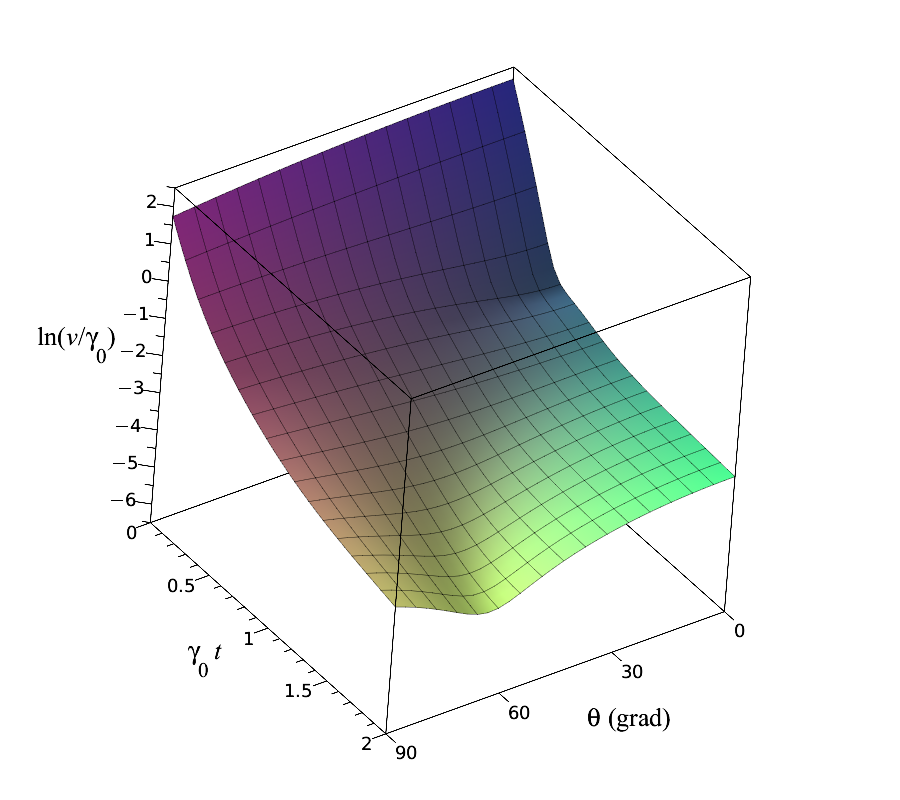}
    \caption{$n_T=0.3,\:\theta_{\Gamma}=0$}
    \label{subfig:2f}
  \end{subfigure}
\caption{
  Speed of evolution
  in the $\gamma_0t$--$\theta$ plane
  computed
for the polarization qubit states~\eqref{eq:pol-qubit} with $\phi=0$
at various values of $n_T$ and $\theta_{\Gamma}$:
(a)~$n_T=0$ and $\theta_{\Gamma}=\pi/2$;
(b)~$n_T=0.3$ and $\theta_{\Gamma}=\pi/2$;
(c)~$n_T=0$ and $\theta_{\Gamma}=\pi/4$;
(d)~$n_T=0.3$ and $\theta_{\Gamma}=\pi/4$;
(e)~$n_T=0$ and $\theta_{\Gamma}=0$;
(f)~$n_T=0.3$ and $\theta_{\Gamma}=0$.
The frequency and relaxation vectors are
  $\bs{\omega}=0.9\gamma_0(0,0,1)$ and $\bs{\gamma}=0.9\gamma_0(\sin\theta_{\Gamma},0,
  \cos\theta_{\Gamma})$, respectively.
  } 
\label{fig:t-theta}
\end{figure*}

The time-dependent part of
Eqs.~\eqref{eq:v0-2} and~\eqref{eq:v2-nt}
is determined by
the zero-temperature density
matrix $\hat{\rho}_0(t)=e^{\hat{\mathcal L}_0 t}\rho(0)$.
For the initial state~\eqref{eq:rho_0-0},
the expression for this matrix 
\begin{align}
  &
    \label{eq:rho0-vs-t}
    \hat{\rho}_0(t)=
    (\hat{\mathcal{K}}^{(+)}_{R}+r)
    \ket{\vc{0}}\bra{\vc{0}}, 
  \\
  &
  \label{eq:rho_0-1}    
    R =P(t)R_0\hcnj{P}(t),
    \:
    r=1-\Tr R
\end{align}
can be deduced with the help of
the identities~\eqref{eq:K0-vac}~--\eqref{eq:KA-Ld-1}
derived in Appendix~\ref{appder}.

The derivative
of $\hat{\rho}_0$ with respect time
is given by
\begin{align}
  \label{eq:v0-1}
  \hat{v}_0\equiv \frac{\partial\hat{\rho}_0}{\partial t}=
    (\hat{\mathcal{K}}^{(+)}_{\dot{R}}+\dot{r})
    \ket{\vc{0}}\bra{\vc{0}},
\end{align}
where dot stands for derivative with respect time,
and
it is not difficult to express the zero-temperature speed of evolution
in terms of the matrix $R$ as follows
\begin{align}
  \label{eq:v0-R-r}
  &
    v_0^2=\Tr\hat{v}_0^2=\Tr \dot{R}^2+\dot{r}^2,
\end{align}
where
\begin{subequations}
  \label{eq:dRR1}
\begin{align}
  &
    \label{eq:dR}
    \dot{r}=-\Tr\dot{R},
    \quad
    \dot{R}=
    P(t)R_1\hcnj{P}(t),
  \\
  &
    \label{eq:R1}
    R_1=L R_0 + R_0 \hcnj{L}=
    -[\gamma_0+(\bs{\gamma},\vc{n})]\sigma_0
    \notag
  \\
  &
    +
    (\bs{\omega}\times\vc{n}-\bs{\gamma}-\gamma_0\vc{n},\bs{\sigma}).
\end{align}
\end{subequations}

\begin{figure*}[!htp]
  \centering
  \begin{subfigure}{.3\textwidth}
    \centering
    \includegraphics[width=1\linewidth]{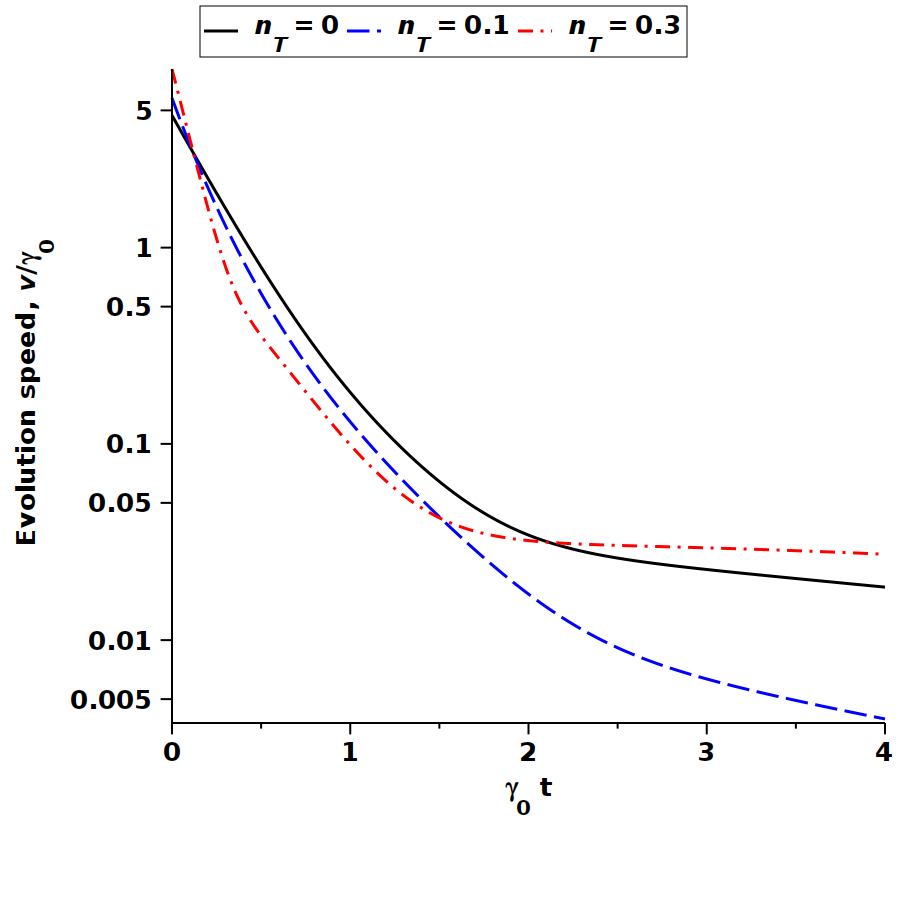}
    \caption{$\omega=0$}
    \label{subfig:wlg}
  \end{subfigure}
  \begin{subfigure}{.3\textwidth}
    \centering
    \includegraphics[width=1\linewidth]{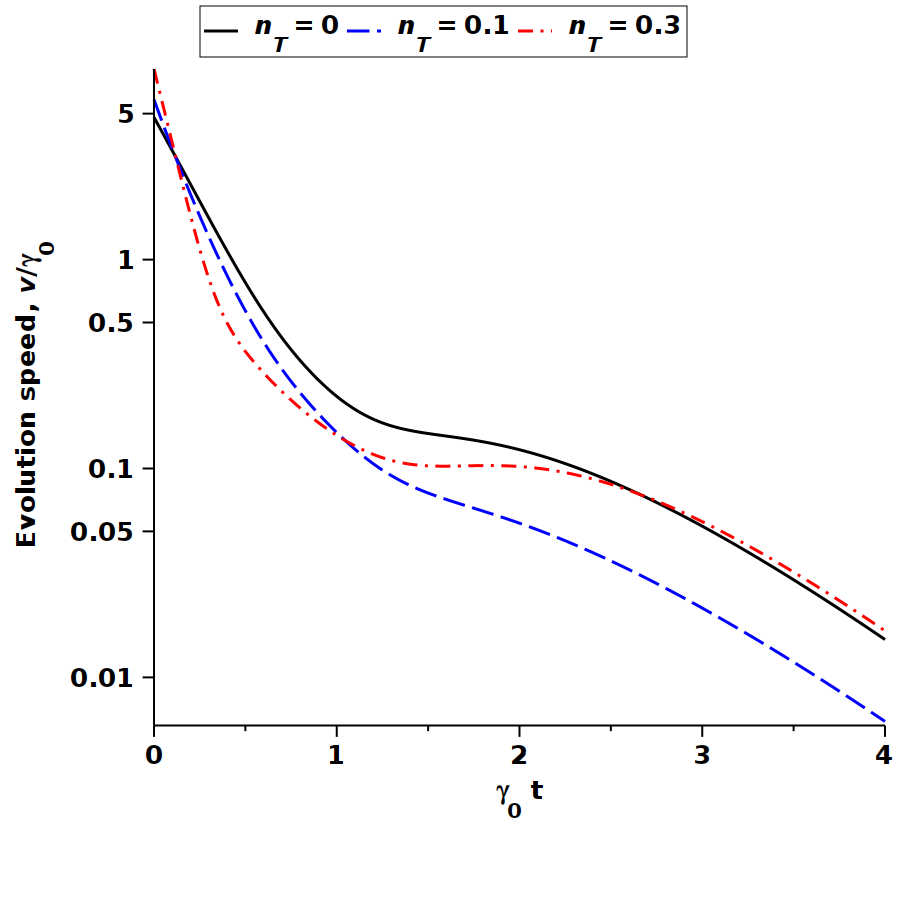}
    \caption{$\omega=\gamma$}
    \label{subfig:weg}
  \end{subfigure}
   \begin{subfigure}{.3\textwidth}
    \centering
    \includegraphics[width=1\linewidth]{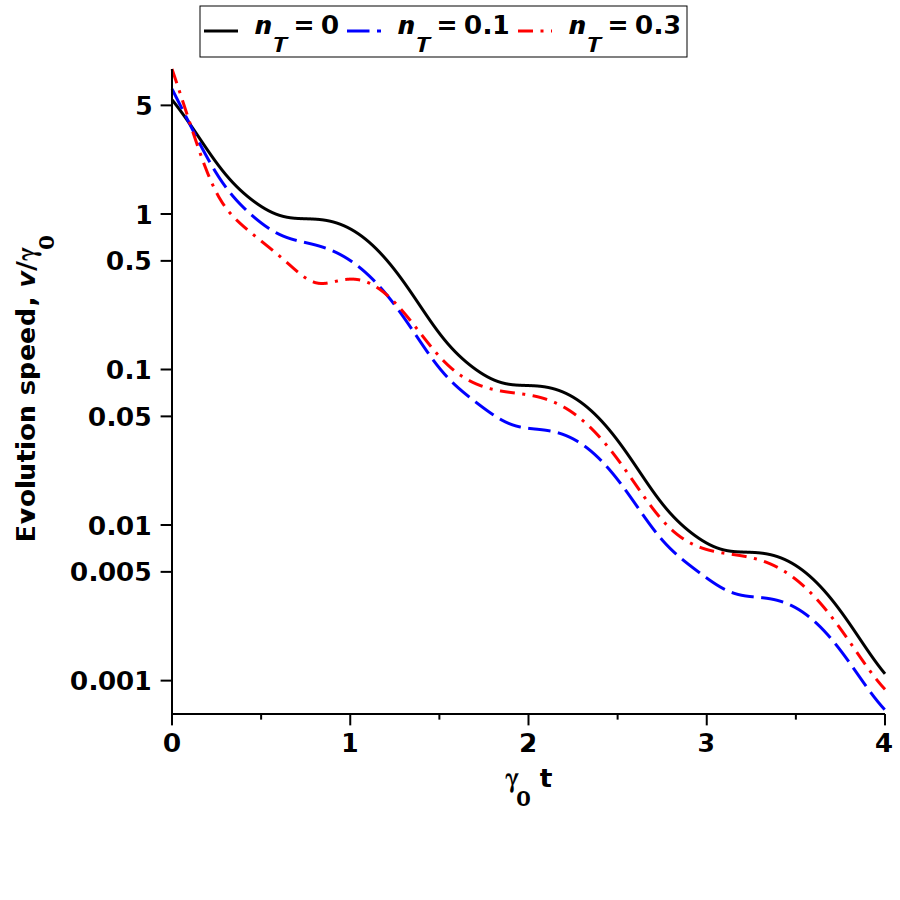}
    \caption{$\omega=3\gamma$}
    \label{subfig:wgg}
  \end{subfigure}
\caption{
  Time dependence of evolution speed
  for the polarization qubit state~\eqref{eq:pol-qubit}
  with $\theta=\pi/4$ and $\phi=0$.
  The frequency and relaxation vectors are
  $\bs{\omega}=(0,0,\omega)$ and $\bs{\gamma}=(0.9\gamma_0,0,0)$, respectively.
  Three cases are shown:
  (a)~$\omega=0$, (b)~$\omega=\gamma$
  and (c) $\omega=3\gamma$.
  Solid, dashed and dotted-dashed lines are evaluated
  at $n_T=0$, $n_T=0.1$ and $n_T=0.3$, respectively.
} 
\label{fig:EP}
\end{figure*}

The latter can be calculated using the identities~\eqref{eq:K0-vac} and
the commutation relations~\eqref{eq:alg-comm-rels}.
The result reads
\begin{align}
  \label{eq:rho1-2}
  \hat{\rho}_1(t)=
(\hat{\mathcal{K}}^{(+)}_{Q}\hat{\mathcal{K}}^{(+)}_{R}+\hat{\mathcal{K}}^{(+)}_{V}+v)
    \ket{\vc{0}}\bra{\vc{0}},
\end{align}
where
\begin{subequations}
  \label{eq:Vv}
\begin{align}
  &
    \label{eq:V}
    V=Q-Q\Tr R - R\Tr Q-\{R,Q\},
  \\
  &
    \label{eq:v}
    v=\Tr(RQ)+\Tr R\Tr Q-\Tr Q.
\end{align}
\end{subequations}


Now it is rather straightforward to obtain
the expression for the first-order correction operator,
$\hat{v}_1$, that enters the evolution speed~\eqref{eq:v2-nt}
given by
\begin{align}
  \label{eq:v1-W-R}
  &
    \hat{v}_1\equiv \frac{\partial\hat{\rho}_1}{\partial t}=
    \bigl(\hat{\mathcal{K}}^{(+)}_{\dot{Q}}\hat{\mathcal{K}}^{(+)}_{R}+\hat{\mathcal{K}}^{(+)}_{Q}\hat{\mathcal{K}}^{(+)}_{\dot{R}}
    \notag
  \\
  &
    +\hat{\mathcal{K}}^{(+)}_{\dot{V}}+\dot{v}\bigr)
    \ket{\vc{0}}\bra{\vc{0}},
\end{align}
where
$\dot{V}$ is the time derivative of the matrix~\eqref{eq:V}
that, in addition to $R$ and $Q$,
depends on $\dot{R}$
(see Eq.~\eqref{eq:dRR1}) and
$\dot{Q}=P(t)\Gamma\hcnj{P}(t)$
with $\Gamma=2[\gamma_0\sigma_0+(\bs{\gamma},\bs{\sigma})]$.


These formulas can be conveniently used for a numerical analysis
of the evolution speed that involves
the evaluation of the matrices $P(t)$ as one of the most important steps.
Equation~\eqref{eq:P-2} gives the matrix $P(t)$
expressed in terms of
the frequency and relaxation rate vectors,
$\bs{\omega}$ and $\bs{\gamma}$,
describing the matrices,
$\Omega$ and $\Gamma/2$,
in the basis of the Pauli matrices.
In our subsequent calculations, we
shall utilize
the angular parameterization of these vectors
given by Eq.~\eqref{eq:gamma-omega}
and, without loss of generality,  assume that
the frequency matrix is diagonal: $\bs{\omega}=(0,0,\omega)$.

Note that, from Eq.~\eqref{eq:P-2},
the matrices $R$
(see Eq.~\eqref{eq:rho_0-1})
and $Q$ (see Eq.~\eqref{eq:W})
are both independent of the frequency $\omega_0$.
Therefore, this frequency
has no effect on the evolution speed~\eqref{eq:v2-nt}.

In Figure~\ref{fig:tht_G},  the semi-log plots present
the numerical results for
time dependencies of the dimensionless evolution speed $v/\gamma_0$
computed at
$\omega/\gamma_0=\gamma/\gamma_0=0.9$
with different values of
the mean number of thermal photons $n_T\in\{0,0.1,0.3\}$
and of the angle
$\theta_{\Gamma}\in\{\pi/2,\pi/4,0\}$
between the frequency vector $\vc{n}_{\Omega}=(0,0,1)$
and the relaxation vector $\vc{n}_{\Gamma}=
(\sin\theta_{\Gamma},0,\cos\theta_{\Gamma})$.

Referring to Fig.~\ref{fig:tht_G},
it can be seen that all the curves are approaching zero in
the long time limit, where the density matrix reaches the thermal equilibrium state. 
Another effect is that,
at elevated temperatures with $n_T=0.1$ and $n_T=0.3$,
the initial value of the evolution speed
$v(0)$ increases as compared to the zero-temperature case with $n_T=0$.
At the same time, the rate of the initial evolution speed decay is
accelerated with temperature.

Figure~\ref{fig:t-theta}
shows the evolution speed surfaces
in the $\gamma_0t$--$\theta$ plane
evaluated at different values of temperature and 
angle $\theta_{\Gamma}$.
It is demonstrated that, in addition to the temperature and angle between
the coherent and incoherent coupling vectors,
$\bs{\omega}$ and $\bs{\gamma}$,
the dynamics of evolution speed
depends on the unit vector
$\vc{n}$ that specifies
the initial state
of the polarization qubit~\eqref{eq:rho_0-0}.
More specifically, from Fig.~\ref{fig:t-theta},
it can be seen
that the speed versus time curves appear to be sensitive
to the orientation of the linear polarization axis:
$(\sin\theta,\cos\theta,0)$.

For instance,
Figures~\ref{subfig:2e} and~\ref{subfig:2f} present the results
for the simplest case, where
$\bs{\omega}=\bs{\gamma}\parallel (0,0,1)$
and the matrix
$\hcnj{P}(t)P(t)=\e^{-2\gamma_0t}\diag(\e^{-2\gamma t},\e^{2\gamma t})$
is diagonal.
As shown in Fig.~\ref{subfig:2e},
at $n_T=0$ and $\theta=0$,
we have
the regime of fast perfectly single-exponential relaxation,
where $v_0\propto \e^{-2(\gamma_0+\gamma)t}$.
This regime changes with $\theta$ and
the relaxation slows down  due to growing contribution
of the terms proportional to $\e^{-2(\gamma_0-\gamma)t}$.
Referring to Fig.~\ref{subfig:2f},
at non-zero temperature,
the dynamics is complicated by
additional products of
fast and slow exponential functions coming
from the first-order correction~\eqref{eq:v1-W-R}.
Similar remarks apply to
the surfaces~\ref{subfig:2a}--\ref{subfig:2d}
with nonvanishing angle between
$\bs{\omega}$ and $\bs{\gamma}$. 

Figure~\ref{fig:EP}
shows
the speed of evolution
computed as a function of the time parameter, $\gamma_0t$,
at different values of the frequency, $\omega$,
assuming that
the dynamical coupling (frequency) vector
$\bs{\omega}=(0,0,\omega)$ is normal to
the incoherent coupling (relaxation) vector
$\bs{\gamma}=(\gamma,0,0)$.
As illustrated in Fig.~\ref{fig:EP},
the exceptional point that takes place at
$\omega=\gamma$ (see Fig.~\ref{subfig:weg})
separates two qualitatively different dynamical regimes.
These regimes are
determined by the parameter $q$ (see Eq.~\eqref{eq:omega-gamma})
that enters the matrix $P(t)$ (see Eq.~\eqref{eq:P-2}):
(a)~the exponential regime (see Fig.~\ref{subfig:wlg})
corresponds to the case, where $\omega$ is smaller than $\gamma$
and the parameter $q$ is real;
(b)~the oscillatory regime (see Fig.~\ref{subfig:wgg})
occurs when $\omega$ exceeds $\gamma$
so that the value of $q$ is imaginary.

\section{Conclusions and discussion}
\label{sec:conclusion}

In this paper, we have studied the Lindblad dynamics of
multimode bosonic systems
governed by the GKSL equation of the form~\eqref{eq:L-rho}
with relations~\eqref{eq:Gamma_pm} describing systems
interacting with the thermal environment.
This thermal bath case involves both the coherent (dynamical) and incoherent (environment-mediated)
intermode couplings represented by the frequency and relaxation matrices:
$\Omega$ and $\Gamma$, respectively.
At the heart of our general approach to dynamics is the Lie algebra of
quadratic combinations of the left and right
superoperators (see Eqs.~\eqref{eq:Nm_nm}--\eqref{eq:Kpm_nm}),
$\{\overleftarrow{\hat{a}^\dagger_i}, \overleftarrow{\hat{a}_i}\}_{i=1}^N$
and
$\{\overrightarrow{\hat{a}^\dagger_i}, \overrightarrow{\hat{a}_i}\}_{i=1}^N$,
with the commutation relations given by Eq.~\eqref{eq:alg-comm-rels-0}.

We have found that
introducing the superoperators associated with matrices
(see Sec.~\ref{subsec:super-assoc-with} and the commutation relations~\eqref{eq:alg-comm-rels})
leads to a natural representation of the algebraic structure of the Liouvillian
(see Eq.~\eqref{eq:L-gen-form})
and significantly simplifies analysis of this algebra. 
This step can be regarded as an extension of the well-known
Jordan-Schwinger approach outlined in Sec.~\ref{subsec:Jordan}
to the case of superoperators.

We have derived the algebraic identities~\eqref{eq:transform}
which are utilized
for computing the differently ordered
jump-eliminating transformations
(see Eqs.~\eqref{eq:T_pm-matr} and~\eqref{eq:T_mp-matr}).
that recast the Liouvillian~\eqref{eq:L-gen-form} into the diagonalized
form~\eqref{eq:L-diag}, where contributions coming from the jump
superoperators (recycling terms) are eliminated.
These transformations are expressed in terms of solutions
to the algebraic Ricatti equations which are analyzed in the
Appendix~\ref{app:Riccati}.
It turns out that these solutions essentially depend on
the matrix~\eqref{eq:Ap-sol} closely related to the algebraic
structure of the steady state statistical operator
(see Eq.~\eqref{eq:rho_ss}).

As a result,
in the limiting case of thermal bath Liouvillians~\eqref{eq:Gamma_pm},
the jump-eliminating
transformations take the simplified form
of the superoperators
given by Eqs.~\eqref{eq:Tpm}
and~\eqref{eq:Tmp}
with the matrices proportional to unity matrix.
In this case,
the transformed Liouvillian, $\hat{\mathcal{L}}$,
and its conjugate, $\hat{\mathcal{L}}^{\sharp}$,
defined using relation~\eqref{eq:L_cnj}
are given by formulas~\eqref{eq:Ld}
and~\eqref{eq:Ld-sharp}, respectively.

Our general result is that
the diagonalized Liouvillian~\eqref{eq:LdHeff} is completely determined
by the effective non-Hermitian Hamiltonian~\eqref{eq:Heff},
$\hat{H}_{\ind{eff}}$,
associated with the matrix $H=\Omega-i\Gamma$
through the Jordan mapping~\eqref{eq:Jordan}.
In contrast to the approximate semiclassical Hamiltonian~\eqref{eq:Heff_0},
the effective Hamiltonian is temperature independent,
and the associated matrix
$H$ governs both the spectral properties of the Liouvillian and
Lindblad dynamics regimes.
In particular, at Liouvillian exceptional points,
this matrix cannot be diagonalized.

As an immediate consequence of the jump-eliminating procedure,
the spectral problem for the Liouvillian
and its conjugate can be analytically treated by expressing
the spectrum and eigenoperators in terms of
the eigenvalues and the eigenstates of the operator
proportional to the Hamiltonian, $\hat{L}_{\ind{eff}}=-i \hat{H}_{\ind{eff}}$
(see Eqs.~\eqref{eq:L-lambda}--\eqref{eq:q-lambda}).
Efficiency of the method is demonstrated
by solving the spectral problem
(see Eqs.~\eqref{eq:1D-spec-d}--~\eqref{eq:unity-1D})
and deriving the eigenmode decomposition
of the matrix density evolved in time
(see Eq.~\eqref{eq:evol-1D})
for the single-mode Liouvillian closely related to the case of
noninteracting modes.
Note that
in Appendix~\ref{app:biorth},
we have employed
the method of generating operators
as a tool that considerably simplifies
the derivation of the biorthogonality relations. 

For the case where the matrix $H$ is diagonalizable,
we have shown how to diagonalize
the effective Hamiltonian, solve the spectral problem,
and obtain the eigenmode expansion for $\hat{\rho}(t)$
(see Eqs.~\eqref{eq:ident-Hpm-Vpm}--\eqref{eq:qmn-N}).
We have also derived an alternative representation
for $\hat{\rho}(t)$ that remains valid even if the matrix $H$
is non-diagonalizable (see Eqs.~\eqref{eq:evol-Pimn}--\eqref{eq:qmn-N}).
The latter is the case where
a Liouvillian exceptional point comes into play.

In addition to the above spectral analysis,
we have shown how our algebraic technique
applied to bimodal bosonic systems
can be used to study
the intermode coupling-induced regimes
of the Lindblad dynamics of photonic polarization modes
and the related problem of the speed of evolution.
In this case, the matrices $\Omega$ and $\Gamma$
expressed in terms of the Pauli matrices
are conveniently parameterized by
the frequency and relaxation vectors,
$\bs{\omega}$ and $\bs{\gamma}$
(see Eq.~\eqref{eq:Omega-Gamma}).

The explicit expression for the matrix exponential
$P(t)=e^{-i Ht}$ given by Eq.~\eqref{eq:P-2}
shows that the dynamical regimes
are governed by the parameter $q$
(see Eq.~\eqref{eq:omega-gamma})
giving
the eigenvalues of the matrix $i H$:
$\lambda_{\pm}=i\omega_0+\gamma_0 \pm q$.
The formula~\eqref{eq:P-2}
is used to deduce two relations~\eqref{eq:EP-cond} describing
the geometry of $\bs{\omega}$ and $\bs{\gamma}$ at
EPs, where the parameter $q$ vanishes along with the difference
between the above eigenvalues:
$|\bs{\omega}|\equiv\omega=|\bs{\gamma}|\equiv\gamma$
and $\bs{\omega}\perp \bs{\gamma}$.

When Eq.~\eqref{eq:P-2} is compared with
its EP limit~\eqref{eq:P-EP},
it is apparent that EPs would manifest themselves
in the effect of EP induced slowdown.
In the special case where
$\bs{\omega}$ is aligned perpendicular to $\bs{\gamma}$,
EP represents the point separating
the regions with  $\omega>\gamma$ (the value of $q$ is imaginary)
and $\omega<\gamma$ ($q$ is real-valued).
So, we have two different dynamical regimes separated by the EP.
A similar result was reported in Ref.~\cite{brody2019evolution} 
for a simple model represented by the qubit system
with the Hamiltonian and the jump operator
taken to be proportional to the Pauli matrices
$\sigma_1$ and $\sigma_3$, respectively.
Note that this qubit model can be transformed into
the spin-1/2 model studied in Ref.~\cite{Minganti:pra:2020}
where the above matrices $\sigma_1$ and $\sigma_3$
are interchanged.
A more complicated example of
a two-level system with EPs
and jump operators
describing three different decoherence channels
was recently reported in~\cite{Xu:njp:2022}.

In order to study the speed of evolution
in the low temperature regime,
where the mean number of thermal photons
$n_T$ is small,
we have deduced the approximate expression for
the superpropagator,
$e^{\hat{\mathcal{L}}t}$,
which linearly depend on $n_T$
(see Eqs.~\eqref{linearres}--~\eqref{eq:W}).
The resulting approximate density matrix~\eqref{eq:rho1-1}
and the speed of evolution~\eqref{eq:v2-nt}
are utilized to perform a numerical analysis
for
the single-photon initial state~\eqref{eq:pol-qubit}
with the density matrix of
the polarization qubit~\eqref{eq:rho_0-0}.
(At $\phi=0$, the qubit is
linearly polarized with the polarization azimuth $\theta$.)

In order to evaluate the evolution speed~\eqref{eq:v2-nt},
we have derived analytical formulas for
the zero-temperature density matrix~\eqref{eq:rho0-vs-t}
and the first-order correction~\eqref{eq:rho1-2}
defined with the help of superoperators
$\hat{\mathcal{K}}^{(+)}_{mn}$ associated with
the time-dependent matrices
$W$ (see Eq.~\eqref{eq:W}),
$R$ (see Eq.~\eqref{eq:rho_0-1}) and $V$ (see Eq.~\eqref{eq:V}). 
Interestingly, these formulas can be used to
demonstrate the difference between the polarization qubit dynamics
and the temporal evolution of
two-level atomic systems
that interact with a bosonic bath
and the jump operators are proportional
to $\sigma_{\pm}=(\sigma_1\pm i\sigma_2)/2$
(see, for instance, equation~(2.19) in Ref.~\cite{Nakazato:pra:2006}).

In addition, our general results for
zero-temperature dynamics of Fock states
provide a useful tool for
the analysis of generalized 
models of passive parity-time-symmetric
optical directional couplers dealing with
two-mode bosonic systems that represent
light traveling along evanescently coupled
waveguides~\cite{Longhi:ol:2020,Hernandez:ol:2023}.

The curves depicted in Fig.~\ref{fig:tht_G}
illustrate the temperature-induced effects in
the dynamics of the evolution speed.
Specifically, 
the initial evolution speed is found to increase with temperature,
and it decays more rapidly at short times as the temperature goes up.
For linearly polarized qubit~\eqref{eq:pol-qubit}
with vanishing $\phi$,
the evolution speed surfaces depicted  in Fig.~\ref{fig:t-theta}
demonstrate sensitivity of the dynamics to
orientation of the polarization axis specified by
the polarization azimuth $\theta$
In Fig.~\ref{fig:EP},
we have also evaluated
the evolution speed
at $\bs{\omega}=(0,0,\omega)$
normal to $\bs{\gamma}=(\gamma,0,0)$
to visualize two qualitatively different dynamical regimes
with $\omega<\gamma$ and $\omega>\gamma$
separated by the EP at $\omega=\gamma$.

The peculiarity of the low-temperature approximation
with the Liouvillian and the superpropagator
expanded as power series in $n_T$
is that
the corresponding approximate superoperators
acting on density matrices with finite-dimensional support
in the Fock basis will produce the operators whose support
is also finite-dimensional.
Thus, by using this approximation,
theoretical treatment of the Lindblad dynamics of
such states that
exemplify
extremes opposite to the well-known Gaussian states
can be performed within the realm of finite-dimensional models.

Our results are applicable to bosonic systems with an arbitrary number of modes,
so it is rather straightforward to
go beyond the limitations of the two-mode system
by taking into consideration additional modes.
For instance, such generalizations will be useful
when studying the dynamics of light
in systems such as twisted fibers~\cite{vavulin2017robust}
and open resonators with overlapping
modes~\cite{Hackenbroich:pra:2003}.
This framework could
also be useful to study measurement-induced phase transitions
in bosons~\cite{Minoguchi:SciPostPhys:2022,Yokomizo:arxiv:2023}.
Among a wide range of potential applications,
note that our method provides an efficient tool for
computing multiple-time correlation functions using
the quantum regression theorem
(a recent discussion of related techniques can be found in Refs.~\cite{Blocher:pra:2019,Khan:pra:2022}).

Our final remark is that, according to Appendix~\ref{app:Riccati},
the thermal bath constraints~\eqref{eq:Gamma_pm}
imposed on the relaxation matrices can be relaxed.
This opens up an opportunity to explore consequences
of applicability of Lindblad dynamics modeling to a wide range of processes. 


\section*{Acknowledgements}

ADK  acknowledges support from the Russian Science Foundation (project No. 24-11-00398).

\appendix

\section{Solutions of algebraic Riccati equations and effective Hamiltonian}
\label{app:Riccati} 

For
the superoperator
$\exp(\hat{\mathcal{K}}^{(\nu)}_{A_{\nu}})$
giving the first part of a jump-eliminating transformation
$\hat{\mathcal{T}}$,
the condition $\Gamma_{\nu}'=0$
(see Eq.~\eqref{eq:Gamma_nu-prime})
leads to the algebraic Riccati equation of the form:
\begin{align}
  \label{eq:app-R1}
      \Gamma_{\nu}+\frac{i}{2}[\Omega,A_{\nu}]-
  \frac{\nu}{2}\{\Gamma_0,A_{\nu}\}+A_{\nu}\Gamma_{-\nu}A_{\nu}=0,
\end{align}
whose solution $A_{\nu}$ determines the matrices
associated with the effective Hamiltonian
(see Eq.~\eqref{eq:OmG0-prime})
\begin{subequations}
    \label{eq:Omg-G0-prime}
\begin{align}
    &
    \label{eq:Omg-prime-2}
    \Omega'= \Omega+i [\Gamma_{-\nu},A_{\nu}],
  \\
  &
    \label{eq:G0-prime-2}
    \Gamma_{0}' = \Gamma_0-\nu\{\Gamma_{-\nu},A_{\nu}\}.
\end{align}
\end{subequations}
Since $\Gamma_0=-(\Gamma_{+}+\Gamma_{-})$,
it is not difficult to see that formula
\begin{align}
  \label{eq:app-Anu-I}
  A_{\nu}=-\nu I
\end{align}
gives a solution of Eq.~\eqref{eq:app-R1}
and the corresponding matrices of
the diagonalized Liouvillian~\eqref{eq:L-diag}
are given by
\begin{align}
    &
    \label{eq:Omg-G0-prime-1}
    \Omega'= \Omega,\quad
    \Gamma_{0}' = \Gamma_0+2 \Gamma_{-\nu}=\nu\Gamma,
\end{align}
where $\Gamma=\Gamma_{-}-\Gamma_{+}>0$. 

At $\nu=-1$,
$\Gamma_0'<0$ and
we have
\begin{align}
  &
      \label{eq:app-Leff}
    \hat{\mathcal L}_d=
    \overleftarrow{\hat{L}_{\mathrm{eff}}}+\overrightarrow{\hat{L}_{\mathrm{eff}}^{\dagger}},
  \quad
     \hat{L}_{\mathrm{eff}}=\hat{J}_{L},\quad 
    L=-i\Omega-\Gamma.
\end{align}


For other solutions taken in the form
\begin{align}
  \label{eq:app-Anu}
  A_{\nu}=X_\nu-\nu I
\end{align}
we obtain equation
\begin{align}
  \label{eq:Xnu-eq}
  \frac{i}{2}[\Omega,X_{\nu}]-
  \frac{1}{2}\{\Gamma,X_{\nu}\}+X_{\nu}\Gamma_{-\nu}X_{\nu}=0
\end{align}
that allows the matrix $X_{\nu}$
to be expressed in terms of the solution
to the linear matrix equation
\begin{align}
  &
    \label{eq:Wnu-eq}
    L W_{\nu} + W_{\nu}\hcnj{L}+ 2\Gamma_{\nu}=0
\end{align}
as follows
\begin{align}
  &
  \label{eq:Xnu}
    X_{\nu}=(W_{-\nu})^{-1}.
\end{align}

After introducing the time dependent matrices
\begin{align}
  &
    \label{eq:Wnu-t}
    W_{\nu}(t)=P(t)W_{\nu}\hcnj{P}(t),
  \\
  &
    \label{eq:Gamma-nu-t}
    \Gamma_{\nu}(t)=P(t)\Gamma_{\nu}\hcnj{P}(t),
    \quad P(t)=e^{L t},
\end{align}
algebraic equation~\eqref{eq:Wnu-eq}
can be cast into the form of the following differential eqution 
\begin{align}
  \label{eq:Wnu-diff-eq}
  \frac{d}{d t} W_{\nu}(t)+ 2 \Gamma_{\nu}(t)=0.
\end{align}
Since $P(\infty)=0$,
we have the boundary condition
$W_{\nu}(\infty)=0$
and the corresponding solution
\begin{align}
  &
  \label{eq:Wnu_t}
  W_{\nu}(t)=-2\int_0^t\Gamma_{\nu}(\tau)\dd
  \tau+2\int_0^{\infty}\Gamma_{\nu}(\tau)\dd \tau
\end{align}
gives the matrix $W_{\nu}$
in the integral form
\begin{align}
  &
    \label{eq:Wnu-sol}
    W_{\nu}=W_{\nu}(0)=2\int_0^{\infty}P(\tau)\Gamma_\nu\hcnj{P}(\tau)\dd \tau.
\end{align}
This result can now be combined with Eq.~\eqref{eq:Xnu}
to yield the following relations for the matrix $A_{\nu}$:
\begin{align}
  &
  \label{eq:app-Anu-2}
    A_{\nu}=(W_{-\nu})^{-1}-\nu I,
  \\
  &
    \label{eq:app-Amnu}
    A_{-\nu}=W_{\nu}^{-1}+\nu I=-A_{\nu}^{-1},
\end{align}
where the identity~\eqref{eq:app-Amnu}
can be deduced using the relation
\begin{align}
  &
    \label{eq:app-Wnu-rel1}
    W_{\nu}=-\nu I + W_{-\nu}
\end{align}
that follows from the equality
\begin{align}
  &
    \label{eq:app-Wnu-rel2}
    2\int_0^{\infty}P(\tau)\Gamma\hcnj{P}(\tau)\dd \tau=I.
\end{align}

We can now
utilize Eq.~\eqref{eq:app-Anu-2} to
evaluate the matrices~\eqref{eq:Omg-G0-prime}.
Our first step is to use 
Eq.~\eqref{eq:Wnu-eq}
for computing the (anti)commutators
\begin{align}
  &
    \label{eq:anticomm-app}
    \{\Gamma_{\nu},W_{\nu}^{-1}\}=
    -\frac{1}{2}
    \{L
    W_{\nu}+W_{\nu}\hcnj{L},W_{\nu}^{-1}\}
    \notag
  \\
  &
    =\Gamma+W_{\nu}^{-1}\Gamma W_{\nu},
  \\
  &
    \label{eq:comm-app}
    i[\Gamma_{\nu},W_{\nu}^{-1}]
    =-\Omega+W_{\nu}^{-1}\Omega W_{\nu}.
\end{align}
The final result
\begin{align}
  &
  \label{eq:app-Gamma-p}
    \Gamma_0'=\Gamma_0-\nu\{\Gamma_{-\nu},(W_{-\nu})^{-1}-\nu I\}
    \notag
  \\
  &
    =
    \nu\Gamma-\nu\{\Gamma_{-\nu},(W_{-\nu})^{-1}\}=
    -\nu (W_{-\nu})^{-1}\Gamma W_{-\nu},
  \\
  &
    \Omega'=\Omega+
    i[\Gamma_{-\nu},(W_{-\nu})^{-1}]
    =(W_{-\nu})^{-1}\Omega W_{-\nu}
\end{align}
suggests that, at $\nu=1$,
the effective Hamiltonian
is given by
\begin{align}
  &
      \label{eq:app-Leff-2}
     \hat{L}_{\mathrm{eff}}=\hat{J}_{\tilde{L}},\quad 
    \tilde{L}= (W_{-})^{-1} L W_{-}.
\end{align}
Note that,
from algebraic identity~\eqref{eq:Jordan-ident-1},
this Hamiltonian and the Hamiltonian
given by Eq.~\eqref{eq:app-Leff}
are related by the similarity
transformation
\begin{align}
  \label{eq:app-Leff12}
  \hat{J}_{\tilde{L}}=e^{-\hat{J}_{V_{-}}}\hat{J}_{L}e^{\hat{J}_{V_{-}}},
  \quad
  e^{V_{-}}=W_{-}
\end{align}
and, thus, their spectral properties
are equivalent.

Our concluding remark concerns the thermal bath Lindlandians
with the relaxation matrices~\eqref{eq:Gamma_pm}.
In this case, identity~\eqref{eq:app-Wnu-rel2}
shows that the matrices $W_{\nu}$
are proportional to the unity matrix
\begin{align}
  \label{eq:app-Wnu-thermal}
  W_{\nu}=\gamma_{\nu} I
\end{align}
and the spectrally equivalent effective Hamiltonians
$\hat{J}_{{L}}$
(see Eq.~\eqref{eq:app-Leff})
and
$\hat{J}_{\tilde{L}}$
(see Eq.~\eqref{eq:app-Leff-2})
are identical:
$\hat{J}_{L}=\hat{J}_{\tilde{L}}$.

\section{Generating operators and biorthogonality relations}
\label{app:biorth}

In this Appendix, we
derive the biorthogonality relations
for
the eigenoperators
of the single-mode Liouvillian $\hat{\mathcal L}$
and its adjoint $\hat{\mathcal L}^{\sharp}$:
$\hat{\rho}_{mn}$
and $\hat{\sigma}_{nm}$
given by Eqs.~\eqref{eq:rho-mk-1D}
and~\eqref{eq:sigma-1D}, respectively.

To this end, we introduce the generating operators
\begin{align}
  &
  \label{eq:gen-op-rho}
    \hat{\rho}(\alpha,\alpha^{\ast})=
    \sum_{m,n=0}^{\infty}\frac{\hat{\rho}_{mn}}{\sqrt{m!n!}}\alpha^{m}(-\alpha^{\ast})^{n}
    \notag
  \\
  &
    =e^{Z|\alpha|^2}e^{\alpha\hat{a}^{\dagger}}\hat{\rho}_0e^{-\alpha^{\ast}\hat{a}},
  \\
  &
     \label{eq:gen-op-sigma}
    \hat{\sigma}(\beta,\beta^{\ast})=
    \sum_{m,n=0}^{\infty}\frac{\hat{\sigma}_{nm}}{\sqrt{n!m!}}\beta^{n}(-\beta^{\ast})^{m}
    \notag
  \\
  &
    =e^{n_T|\beta|^2}e^{\beta\hat{a}^{\dagger}} e^{-\beta^{\ast}\hat{a}}
\end{align}
that can be computed using
the explicit expressions for $\hat{\rho}_{mn}$
and $\hat{\sigma}_{nm}$
(see Eqs.~\eqref{eq:rho-mk-1D}
and~\eqref{eq:sigma-1D}).

Our next step is
to use the identity
\begin{align}
  \label{eq:tr-nm-0}
  \mathrm{Tr} [(\hat{a}^{\dagger})^n\hat{\rho}_0(\hat{a})^m]=
  n! Z^{n+1} \delta_{nm},
\end{align}
that can be obtained
by integrating the Husimi symbol
(the $Q$-symbol)
of the normally ordered operator
(see normally ordered expression for $\hat{\rho}_0$ given in Eq.~\eqref{eq:rho_0-1D}),
for derivation of the following relation:
\begin{align}
  \label{eq:tr-nm}
  \mathrm{Tr} [e^{\alpha\hat{a}^{\dagger}}\hat{\rho}_0e^{-\alpha^{\ast}\hat{a}}]=
  Z e^{-Z|\alpha|^2}.
\end{align}
This relation can now be combined 
with the formula
\begin{align}
  \label{eq:anti-norm}
  e^{\beta\hat{a}^{\dagger}} e^{-\beta^{\ast}\hat{a}}=e^{|\beta|^2}e^{-\beta^{\ast}\hat{a}} e^{\beta\hat{a}^{\dagger}} 
\end{align}
linking normally and antinormally ordered products
of the exponentials that enter the generating operator~\eqref{eq:gen-op-sigma}
to deduce our key analytical result
\begin{align}
  \label{eq:tr-gen-ops}
  \mathrm{Tr} [\hat{\sigma}(\beta,\beta^{\ast})(\hat{\rho}(\alpha,\alpha^{\ast})]=
  Z e^{-Z(\alpha\beta^{\ast}+\alpha^{\ast}\beta)}.
\end{align}
A comparison between power series expansions
for the left- and right-hand sides of Eq.~\eqref{eq:tr-gen-ops}
immediately leads to the biorthogonality relations of the form:
\begin{align}
  \label{eq:biorth-rel}
  \mathrm{Tr} [\hat{\sigma}_{nm}\hat{\rho}_{m'n'}]=\delta_{nn'}\delta_{mm'}Z^{n+m+1}.
\end{align}

\section{Superoperator of evolution in linear approximation}
\label{appder}

In this Appendix, we shall detail calculations related to
the low-temperature regime where the mean photon number
$n_T$ is small.
In this regime, the Liouvillian and the superpropagator
can be approximated by a sum
of their zero-temperature expressions
and the first-order corrections, which are linear in $n_T$.

In the low-temperature regime,
the Liouvillian $\hat{\mathcal L}_0$
give the zero-order approximation.
Now it is our task to evaluate
the first order corrections that are proportional to $n_T$. 
To this end, we combine formula~\eqref{lo} with Eq.~\eqref{eq:Ld}
to express the Liouvillian $\hat{\mathcal L}$
in terms of $\hat{\mathcal L}_0$ as follows
\begin{align}
  &
  \label{llo}
    \hat{\mathcal L}=e^{\frac{n_T}{n_T+1}\hat{\mathcal{K}}^{(+)}_{I}} e^{-n_T\hat{\mathcal{K}}^{(-)}_{I}}
    \hat{\mathcal L}_0 e^{n_T\hat{\mathcal{K}}^{(-)}_{I}}
    e^{-\frac{n_T}{n_T+1}\hat{\mathcal{K}}^{(+)}_{I}}
    \notag
  \\
  &
  \approx
    \hat{\mathcal L}_0+n_T \hat{\mathcal L}_1,
 \end{align}
where $\hat{\mathcal L}_1=[\hat{\mathcal{K}}^{(+)}_{I}-\hat{\mathcal{K}}^{(-)}_{I},\hat{\mathcal L}_0]$
is the first-order correction that
can be evaluated using the commutation relations~\eqref{eq:alg-comm-rels}.
The result is
\begin{align}
  &
    \label{eq:L1}
    \hat{\mathcal L}_1=\Delta\hat{\mathcal L}_\Gamma=
\hat{\mathcal{K}}_{\Gamma}^{(+)}-2\hat{\mathcal{K}}_{\Gamma}^{(0)}+\hat{\mathcal{K}}_{\Gamma}^{(-)}. 
\end{align}
Similarly, in the linear approximation, the superoperator of evolution
\begin{align}
  &
  \label{eq:evol-1}
    e^{\hat{\mathcal L}t}=e^{\frac{n_T}{n_T+1}\hat{\mathcal{K}}^{(+)}_{I}} e^{-n_T\hat{\mathcal{K}}^{(-)}_{I}}
    e^{\hat{\mathcal L}_0t} e^{n_T\hat{\mathcal{K}}^{(-)}_{I}}
    e^{-\frac{n_T}{n_T+1}\hat{\mathcal{K}}^{(+)}_{I}}
    \notag
  \\
  &
  \approx
    e^{\hat{\mathcal L}_0t}+n_T[\hat{\mathcal{K}}^{(+)}_{I}-\hat{\mathcal{K}}^{(-)}_{I},e^{\hat{\mathcal L}_0 t}]
    \notag
  \\
  &
    =
  \bigl\{\hat{I}+n_T\hat{U}_1(t)\bigr\} e^{\hat{\mathcal L}_0t}
\end{align}
is determined by the operator
\begin{align}
  \label{eq:U1-in}
  \hat{U}_1(t)=
  \hat{\mathcal{K}}^{(+)}_{I}-\hat{\mathcal{K}}^{(-)}_{I}
  -e^{\hat{\mathcal L}_0t}
  (\hat{\mathcal{K}}^{(+)}_{I}-\hat{\mathcal{K}}^{(-)}_{I})e^{-\hat{\mathcal L}_0 t}
\end{align}
describing the first-order correction for the superpropagator $e^{\hat{\mathcal L} t}$.


In order to deduce the expression for $\hat{U}_1(t)$
given by Eq.~\eqref{eq:U1-out},
we cast the time-dependent term
on the right-hand side of Eq.~\eqref{eq:U1-in}
into the form:
\begin{align}
  &
  \label{eq:U1-part}
  e^{\hat{\mathcal L}_0t}(\hat{\mathcal{K}}^{(+)}_{I}-\hat{\mathcal{K}}^{(-)}_{I})e^{-\hat{\mathcal L}_0t} 
  \notag
  \\
  &
    =e^{-\hat{\mathcal{K}}^{(-)}_{I}} e^{\hat{\mathcal L}_dt} e^{\hat{\mathcal{K}}^{(-)}_{I}}
    (\hat{\mathcal{K}}^{(+)}_{I}-\hat{\mathcal{K}}^{(-)}_{I}) e^{-\hat{\mathcal{K}}^{(-)}_{I}}
    e^{-\hat{\mathcal L}_d t} e^{\hat{\mathcal{K}}^{(-)}_{I}}
    \notag
  \\
  &
    =e^{-\hat{\mathcal{K}}^{(-)}_{I} }e^{\hat{\mathcal L}_d t}
    (\hat{\mathcal{K}}^{(+)}_{I}+2\hat{\mathcal{K}}^{(0)}_{I}) e^{-\hat{\mathcal L}_d t} e^{\hat{\mathcal{K}}^{(-)}_{I}},
\end{align}
time dependence is governed by the diagonalized superpropagator
$e^{\hat{\mathcal L}_d t}$.

Our next step utilizes the algebraic identity
(see Eq.~\eqref{eq:ident-Hpm-Vpm} for a similar result
used to diagonalize the effective Hamiltonian)
\begin{align}
  &
  \label{eq:ident-Apm-Vpm}
    e^{\overleftarrow{\hat{V}_{+}}+\overrightarrow{\hat{V}_{-}}}
  \{\overleftarrow{\hat{A}_{+}}\overrightarrow{\hat{A}_{-}}\}
  e^{-\overleftarrow{\hat{V}_{+}}-\overrightarrow{\hat{V}_{-}}}=
  \overleftarrow{\hat{A}_{+}'}\overrightarrow{\hat{A}_{-}'},
  \notag
  \\
  &
    \hat{A}_{\pm}'=e^{\pm\hat{V}_{\pm}}\hat{A}_{\pm}e^{\mp\hat{V}_{\pm}}
\end{align}
with $\hat{V}_{+}=\hat{V}_{-}^{\dagger}=\hat{L}_{\mathrm{eff}}t=\hat{J}_L t$.
By using this identity and Eq.~\eqref{eq:Jordan-ident-2}
we can simplify the operators that enter Eq.~\eqref{eq:U1-part} 
as follows
\begin{align}
  &
    \label{eq:K0-ident-app}
  e^{\hat{\mathcal L}_d t} \hat{\mathcal{K}}^{(0)}_{I} e^{-\hat{\mathcal L}_d t}=\hat{\mathcal{K}}^{(0)}_{I}
  \\
  &
    \label{eq:Kp-ident-app}
    e^{\hat{\mathcal L}_d t} \hat{\mathcal{K}}^{(+)}_{I}
    e^{-\hat{\mathcal L}_d t}=\hat{\mathcal{K}}^{(+)}_{PP^{\dagger}},
\end{align}
where $PP^{\dagger}=e^{L t}e^{L^{\dagger} t}$.
For the two-mode systems, the closed-form expression for the matrix product $PP^{\dagger}$ can
be easily computed from Eq.~\eqref{eq:P-2}.

We can now substitute the relations
\begin{align}
  &
    \label{eq:Kp-rel-app}
  e^{-\hat{\mathcal{K}}^{(-)}_{I}} \hat{\mathcal{K}}^{(+)}_{PP^{\dagger}}
    e^{\hat{\mathcal{K}}^{(-)}_{I}}=\hat{\mathcal{K}}^{(+)}_{PP^{\dagger}}-2\hat{\mathcal{K}}^{(0)}_{PP^{\dagger}}
    \notag
  \\
  &
    +\hat{\mathcal{K}}^{(-)}_{PP^{\dagger}}, 
  \\
  \label{eq:K0-rel-app}
  &
    e^{-\hat{\mathcal{K}}^{(-)}_{I}} \hat{\mathcal{K}}^{(0)}_{I} e^{\hat{\mathcal{K}}^{(-)}_{I}}=\hat{\mathcal{K}}^{(0)}_{I}-\hat{\mathcal{K}}^{(-)}_{I}
\end{align}
along with Eqs.~\eqref{eq:K0-ident-app}-\eqref{eq:Kp-ident-app}
into Eq.~\eqref{eq:U1-part} to deduce
the result leading to Eq.~\eqref{eq:U1-out}.

In order to obtain higher order corrections
and go beyond the linear approximation,
we have to consider the following commutation relations 
\begin{gather}
  [\hat{\mathcal{K}}^{(s)}_{A},e^{\hat{\mathcal L}^{(d)}t}]=
  \big(\hat{\mathcal{K}}^{(s)}_{A}+e^{\hat{\mathcal L}^{(d)}t} \hat{\mathcal{K}}^{(s)}_{A}
  e^{-\hat{\mathcal L}^{(d)}t}
  \big)e^{\hat{\mathcal L}^{(d)}t},
\end{gather}
where $s\in\{0,\pm\}$ and $A$ is an arbitrary matrix.
Similar to Eq.~\eqref{eq:K0-ident-app}--\eqref{eq:Kp-ident-app},
derivation of  the closed-form expressions
for $e^{\hat{\mathcal L}^{(d)}t} \hat{\mathcal{K}}^{(\pm)}_{A} e^{-\hat{\mathcal L}^{(d)}t}$:
\begin{align}
  &
    \label{eq:Kp-gen-app}
  e^{\hat{\mathcal L}_d t} \hat{\mathcal{K}}^{(+)}_{A}  e^{-\hat{\mathcal L}_d t}=
    \hat{\mathcal{K}}^{(+)}_{e^{L t}Ae^{L^{\dagger} t}},
  \\
  &
    \label{eq:Km-gen-app}
  e^{\hat{\mathcal L}_d t} \hat{\mathcal{K}}^{(-)}_{A}  e^{-\hat{\mathcal L}_d t}
  =\hat{\mathcal{K}}^{(-)}_{e^{-L^{\dagger} t}A e^{e^{-L t}}}   
\end{align}
uses Eqs.~\eqref{eq:ident-Apm-Vpm}
and~\eqref{eq:ident-Hpm-Vpm}
in combination
with the identities~\eqref{eq:Jordan-ident-1}--\eqref{eq:Jordan-ident-2}.
Note that Eq.~\eqref{eq:Kp-gen-app} leads to the commutation relation
\begin{align}
    &
    \label{eq:KA-Ld}
      [\hat{\mathcal L}_d, \hat{\mathcal{K}}^{(+)}_{A}] =
    \hat{\mathcal{K}}^{(+)}_{L A+A L^{\dagger}}.
\end{align}

In the conclusion of this Appendix,
we detail algebraic calculations of the zero-temperature
speed of evolution
 for
 the initial density matrix~\eqref{eq:rho_0-0}
 representing the state of the polarization qubit.
These calculations use the identities
\begin{align}
  &
    \label{eq:K0-vac}
    \hat{\mathcal{K}}^{(-)}_{A}\ket{\vc{0}}\bra{\vc{0}}=
  \hat{\mathcal{N}}^{(-)}_{A}\ket{\vc{0}}\bra{\vc{0}}=
    \hat{\mathcal{L}}_{d}\ket{\vc{0}}\bra{\vc{0}}=0,
    \notag
  \\
  &
  2\hat{\mathcal{K}}^{(0)}_{A}\ket{\vc{0}}\bra{\vc{0}}=
  \Tr A \ket{\vc{0}}\bra{\vc{0}}
\end{align}
that can be combined with Eqs.~\eqref{eq:Kp-rel-app},~\eqref{eq:Kp-gen-app} and~\eqref{eq:KA-Ld}
to deduce the relations
\begin{align}
 &
  \label{eq:KA-1}
  e^{\pm \hat{\mathcal{K}}^{(-)}_{I}} \hat{\mathcal{K}}^{(+)}_{A}\ket{\vc{0}}\bra{\vc{0}}
  =
  [\hat{\mathcal{K}}^{(+)}_{A}\pm\Tr A]\ket{\vc{0}}\bra{\vc{0}},
  \\
  &
    \label{eq:KA-eLd}
    e^{\hat{\mathcal L}_d t} \hat{\mathcal{K}}^{(+)}_{A}\ket{\vc{0}}\bra{\vc{0}}=
    \hat{\mathcal{K}}^{(+)}_{P(t)AP^{\dagger}(t)}\ket{\vc{0}}\bra{\vc{0}},
  \\
  &
    \label{eq:KA-Ld-1}
    \hat{\mathcal L}_d \hat{\mathcal{K}}^{(+)}_{P(t)AP^{\dagger}(t)}\ket{\vc{0}}\bra{\vc{0}}=
    \hat{\mathcal{K}}^{(+)}_{P(t)A_1P^{\dagger}(t)}\ket{\vc{0}}\bra{\vc{0}},
\end{align}
where $P(t)=e^{L t}$ and $A_1=L A+A L^{\dagger}$.
It is now rather straightforward to
evaluate
the operator product
$e^{-\hat{\mathcal{K}}^{(-)}_{I}}e^{\hat{\mathcal L}_d t}
e^{\hat{\mathcal{K}}^{(-)}_{I}}\hat{\mathcal{K}}^{(+)}_{R_0}\ket{\vc{0}}\bra{\vc{0}}$
and obtain
the zero-temperature density matrix
$\hat{\rho}_0(t)$ given by Eq.~\eqref{eq:rho0-vs-t}.
\bibliography{bibliography1,quant}

\end{document}